\documentclass[acmtog,authorversion,balance=false]{acmart}

\setcopyright{acmlicensed}\acmConference[SIGGRAPH Conference Papers '25]{Special Interest Group on Computer Graphics and Interactive Techniques Conference Conference Papers }{August 10--14, 2025}{Vancouver, BC, Canada}

\acmBooktitle{Special Interest Group on Computer Graphics and Interactive Techniques Conference Conference Papers (SIGGRAPH Conference Papers '25), August 10--14, 2025, Vancouver, BC, Canada}

\acmDOI{10.1145/3721238.3730713}
\acmISBN{979-8-4007-1540-2/2025/08}

\setcitestyle{nosort,square}

\usepackage{microtype} 
\usepackage[nosort,capitalize]{cleveref}
\usepackage{mathtools}
\usepackage{enumitem}
\usepackage{nicefrac}
\usepackage{bm}
\usepackage{booktabs}
\usepackage{bbm}
\usepackage[normalem]{ulem}
\usepackage{xcolor}
\usepackage{wrapfig}

\definecolor{vecred}{RGB}{240,90,36}

\newcommand{\bgsurf}{\mathcal{M}_\text{b}}

\newcommand{\E}{\mathbb{E}}
\newcommand{\dd}{\,\mathrm{d}}
\newcommand{\dt}{\dd t}
\newcommand{\loss}{\mathcal{L}}

\newcommand{\p}{\mathbf{p}}
\newcommand{\tmax}{t_\mathrm{max}}
\newcommand{\vx}{\mathbf{x}}
\newcommand{\occupancy}{\alpha}
\newcommand{\Lp}{L_\p}
\newcommand{\Lb}{L_\mathbf{b}}
\newcommand{\fb}{f_\mathbf{b}}

\DeclareMathOperator*{\argmin}{argmin}

\definecolor{hlcol}{RGB}{255,200,200}

\newcommand{\attul}[1]{\uuline{\textcolor{black}{#1}}}
\newcommand{\detul}{}

\begin{document}
\fancyhead{}  

\title{Radiance Surfaces: Optimizing Surface Representations with a 5D Radiance Field Loss}

\author{Ziyi Zhang}
\affiliation{
  \institution{École Polytechnique Fédérale de Lausanne (EPFL) and NVIDIA}
  \country{Switzerland}
}
\email{ziyi.zhang@epfl.ch}

\author{Nicolas Roussel}
\affiliation{
  \institution{École Polytechnique Fédérale de Lausanne (EPFL)}
  \city{Lausanne}
  \country{Switzerland}
}
\email{nicolas.roussel@epfl.ch}

\author{Thomas Müller}
\affiliation{
  \institution{NVIDIA}
  \country{Switzerland}
}
\email{tmueller@nvidia.com}

\author{Tizian Zeltner}
\affiliation{
  \institution{NVIDIA}
  \country{Switzerland}
}
\email{tzeltner@nvidia.com}

\author{Merlin Nimier-David}
\affiliation{
  \institution{NVIDIA}
  \country{Switzerland}
}
\email{mnimierdavid@nvidia.com}

\author{Fabrice Rousselle}
\affiliation{
  \institution{NVIDIA}
  \country{Switzerland}
}
\email{frousselle@nvidia.com}

\author{Wenzel Jakob}
\affiliation{
    \institution{École Polytechnique Fédérale de Lausanne (EPFL) and NVIDIA}
    \country{Switzerland}
}
\email{wenzel.jakob@epfl.ch}

\renewcommand{\shortauthors}{Zhang et al.}

\begin{teaserfigure}
    \centering
    \includegraphics[width=\textwidth]{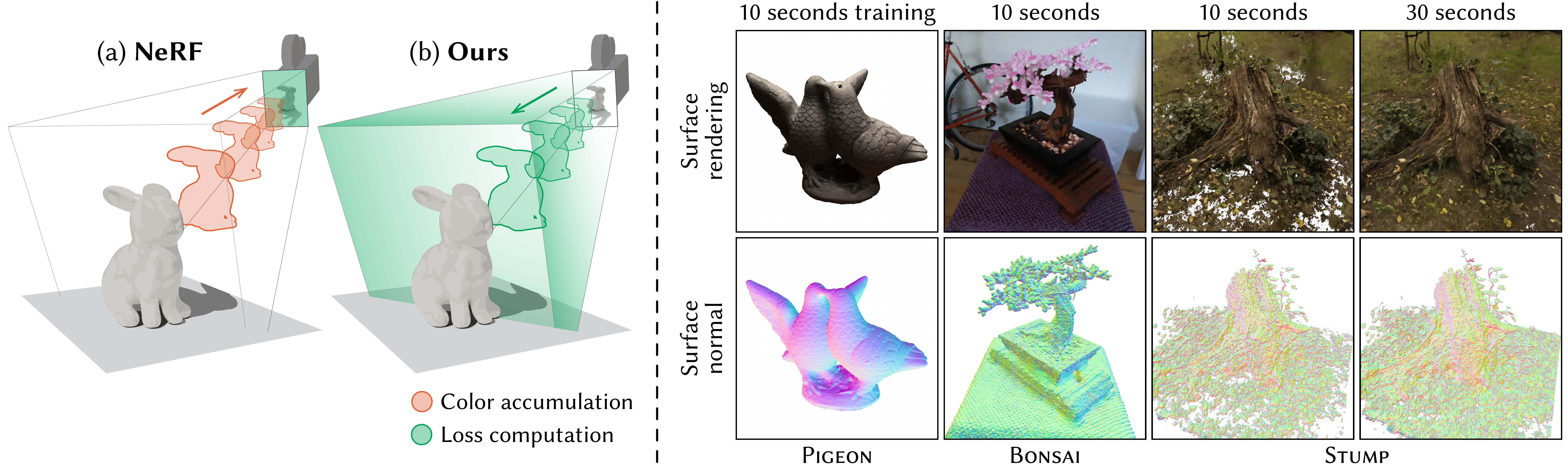}
    \caption{\label{fig:teaser}%
        Our method reconstructs surfaces with the speed and robustness of NeRF-style methods.
        \textbf{Left:} In contrast to volume-based methods that minimize 2D image losses, as shown in (a), we adopt a spatio-directional radiance field loss formulation, as shown in (b).
        At each step, our method considers a distribution of optically independent surfaces, increasing the confidence of candidates that agree with the reference imagery.
        \textbf{Right:} A meaningful surface can be extracted at any iteration during optimization.
    }
\end{teaserfigure}

\begin{abstract}
    We present a fast and simple technique to convert images into a radiance
    surface-based scene representation. Building on existing radiance volume
    reconstruction algorithms, we introduce a subtle yet impactful modification
    of the loss function requiring changes to only a few lines of code:
    instead of integrating the radiance field along rays and supervising the
    resulting images, we project the training images into the scene to directly
    supervise the spatio-directional radiance field.

    The primary outcome of this change is the complete removal of alpha
    blending and ray marching from the image formation model, instead moving
    these steps into the loss computation. In addition to promoting
    convergence to surfaces, this formulation assigns explicit semantic meaning
    to 2D subsets of the radiance field, turning them into well-defined
    radiance surfaces. We finally extract a level set from this representation,
    which results in a high-quality radiance surface model.

    Our method retains much of the speed and quality of the baseline algorithm.
    For instance, a suitably modified variant of Instant~NGP maintains
    comparable computational efficiency, while achieving an average PSNR that
    is only 0.1 dB lower. Most importantly, our method generates explicit
    surfaces in place of an exponential volume, doing so with a level of
    simplicity not seen in prior work.
\end{abstract}

\maketitle


\section{Introduction}
\label{sec:intro}

The task of reconstructing surfaces from a set of photographs has been a long-standing challenge~\cite{ReconstructionSurvey}.
The appeal of surface representations, aside of their natural alignment with the physical reality of objects, lies in their suitability for editing, animation and efficient rendering, which explains their near-ubiquitous use in 3D graphics applications.
Unfortunately, the optimization landscape of a differentiably rendered surface tends to be non-convex and riddled with local minima.
Consequently, the resulting methods are often too fragile to handle complex, real-world scenes.

This problem can be cleverly sidestepped~\citep{mildenhall2020nerf, kerbl20233d} by switching to a volumetric formulation of light transport.
The derivative of a continuous volumetric representation is not only easier to evaluate, but it also leads to a smoother loss landscape that brings enhanced robustness and scalability.
However, these improvements come at the cost of a more involved surface extraction process requiring additional heuristics, such as surface-promoting regularizers~\citep{wang2021neus} or multi-stage optimization~\citep{guedon2024sugar}.

In this work, we seek a simple and direct approach to optimize surfaces that retains the robustness and convergence speed of volumetric methods.
Our proposed method builds on a simple yet powerful idea: optimizing a \emph{distribution over surfaces}.
Concretely, we propose projecting the training photographs into the scene and minimizing the attenuated difference between the resulting light field and the spatial-directional emission originating from the \mbox{surface distribution}.

\begin{figure}[t]
    \centering
    \includegraphics[width=1.0\columnwidth]{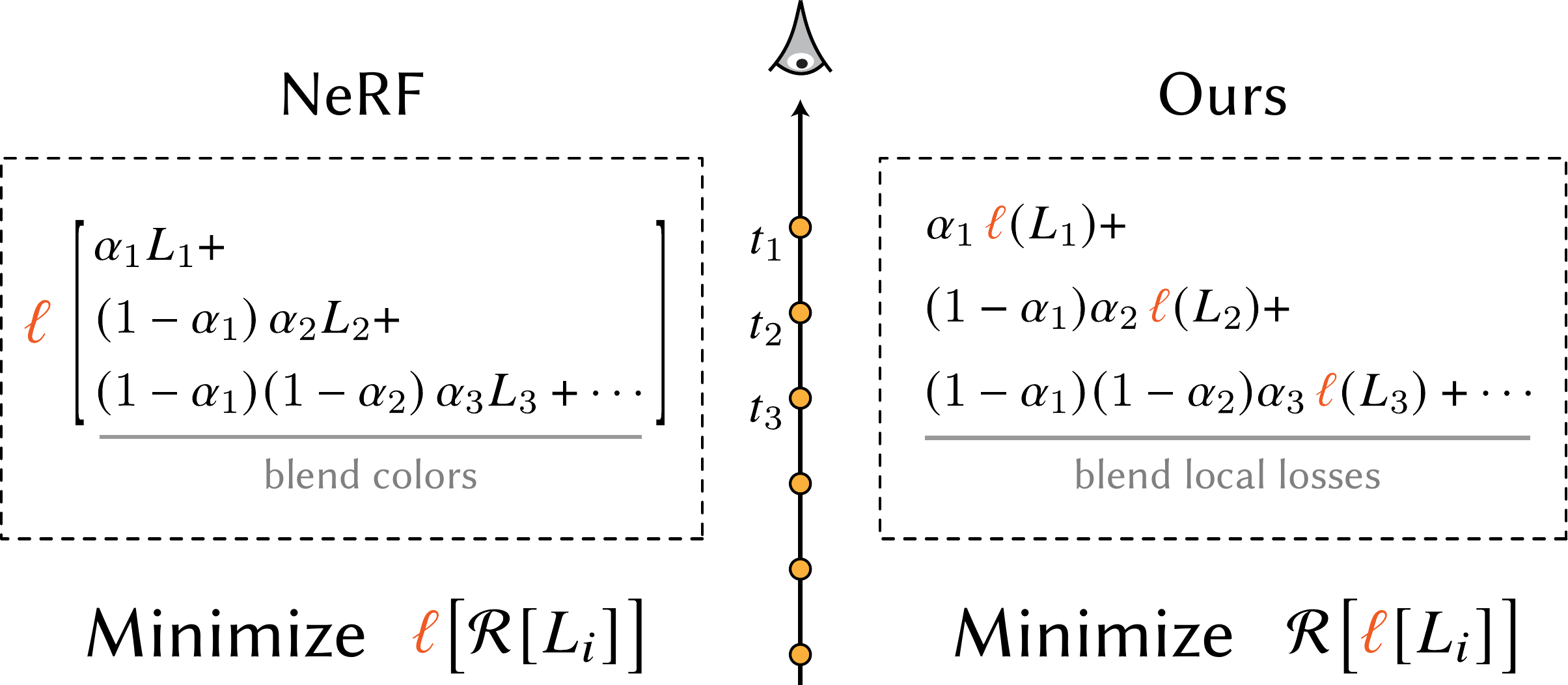}
    \caption{\label{fig:nerf_comparison}%
        \textbf{Comparison of the loss in volumetric optimization and our radiance field loss.}
        We denote alpha blending by $\mathcal{R}$ and the color difference metric as ${\color{vecred}\ell}(L) \coloneq {\color{vecred}\ell}(L, L_\text{target})$ and drop its dependency on the target color for simplicity.
        Traditional volumetric reconstruction minimizes the \emph{image-space loss of blended colors}.
        In contrast, our method minimizes \emph{a blended radiance field loss} that yields a distribution of surfaces out of which a surface representation can be trivially extracted, e.g., via marching cubes.
    }
\end{figure}

The resulting \emph{radiance field loss} considers each point along a ray as a surface candidate, individually optimized to match that ray's pixel color, leading to the desired distribution over surfaces.
One benefit is that points along a ray receive independent gradients, allowing the color or density to simultaneously increase at one point and decrease at another.
This is notably different from the volumetric approach, which integrates the color along the ray prior to the loss computation (see Figure~\ref{fig:teaser}, left).
That is, with volume reconstruction, all points along a ray receive gradients with the same sign if their integrated color is too dark or bright, leading to correlated adjustments.

Interestingly, our proposed radiance field loss gives rise to equations remarkably similar to those of volumetric reconstruction methods (see \autoref{fig:nerf_comparison}).
In practical terms, this means that our method is simple to integrate into existing volumetric frameworks.
It also means that we inherit many advantages of these prior works without having to resort to additional heuristics to extract a surface.
While we have not focused on competing with existing methods in terms of metrics, our proof-of-concept implementation in Instant NGP~\citep{mueller2022instant} consists of only a few modified lines of code in the core algorithm and runs at roughly the same speed (in terms of PSNR vs.\ time) while producing surfaces whose PSNR is, on average, only $0.1$ dB lower than that of the volumetric baseline.


\section{Related work}
\label{sec:related}

This section reviews related work in the field of 3D surface reconstruction for novel view synthesis and tasks centered on geometric representations.
Because this is such an active field, we highlight particularly salient prior works rather than attempting an exhaustive survey.
As such, we only cover differentiable rendering and omit classical techniques like silhouette carving~\citep{Laurentini:Silhouette:1994}.

\paragraph{Evolving a surface}
The first works on differentiable rendering embraced the high-level approach of optimizing an initial guess of a shape via gradient descent~\citep{Loper:OpenDR:2014},
variously representing the surface using SDF level sets~\citep{zhang2021physg,vicini2022differentiable,zichen2024relaxedboundary}, triangle meshes \citep{Nicolet2021Large}, points~\citep{Chen:Dipoles:2024}, or hybrids~\citep{munkberg2022extracting}.
Regardless of the underlying representation, it remains challenging to achieve satisfactory results in this way: this is
partly due to the complex loss
landscape of an evolving surface, and partly due to the numerical difficulties of computing visibility-induced gradients~\citep{Loubet2019Reparameterizing,Zhang:2020:PSDR,Zhang2023Projective}. Without intricate special-case handling, the optimization often fails when topological changes are required~\citep{mehta2023theory}, or when the surface does not overlap with the target shape~\citep{Xing2023EPSM}.
Our work sidesteps these limitations by replacing the surface boundary with a distribution over surfaces.

\paragraph{Extracting geometry from a volume}

After the advent of radiance volume reconstruction (NeRF) for novel view synthesis~\citep{mildenhall2020nerf}, researchers developed various regularizers and parameterizations of radiance volumes to ensure that their level sets yield plausible geometry~\citep{wang2021neus,Yariv2021Volume,yariv2023bakedsdf}.
Surfaces can then be extracted using established algorithms like marching cubes.
However, while efficient NeRF implementations reconstruct in seconds to minutes~\citep{mueller2022instant}, methods in the aforementioned line of work require hours of computation~\citep{li2023neuralangelo} or result in substantially reduced quality~\citep{wang2023neus2}.
In contrast, our method largely preserves the reconstruction speed and quality of the baseline NeRF method.

In real-world reconstruction tasks, it is often ambiguous whether fine details should be attributed to local color variation or geometric features.
The optimal choice depends on whether the intended application emphasizes novel view synthesis performance or reconstruction of smooth surface geometry.
In the former case, our method is a drop-in replacement, e.g., for MobileNeRF~\citep{chen2023mobilenerf}.
For applications requiring smoother geometry, we propose a lightweight Laplacian regularizer that maintains the efficiency of our method, while delivering results comparable to significantly more \mbox{complex algorithms~\citep{Huang2DGS2024,guedon2024sugar}.}

\paragraph{Optimizing a distribution over surfaces}

Several prior works conceptualized volumetric reconstruction as optimizing a distribution over surfaces~\citep{Seyb:2024:Unified,Miller:VOS:2024,wang2021neus}.
These methods however represent objects as the union of multiple interacting surfaces (whose contributions are integrated along the ray), which conflicts with our end goal of extracting a \emph{single} surface to model an object geometry.
Instead, we build upon the ``many worlds'' concept proposed by~\citet{zhang2024many}, which considers a distribution of \emph{non-interacting} surfaces, and apply it to the problem of radiance surface reconstruction.
We show how, in this context, the many worlds concept gives rise to a simple equation dual to the one used in NeRF frameworks; see \autoref{fig:nerf_comparison}.


\section{Method}
\label{sec:method}

In this section, we derive our radiance field loss (\autoref{fig:nerf_comparison}) by progressively transforming the optimization of a single evolving surface.
While the final result resembles volumetric reconstruction, this progression demonstrates that the method's origins are surface-based.

\subsection{Non-local surface perturbation}
\label{sec:many_worlds_perturbation}

Differentiating a rendering with respect to geometry reveals how small geometric perturbations affect the resulting image.
However, because these derivatives are only nonzero on the surfaces themselves, they tend to cause convergence issues when used in optimizations.

To overcome this limitation, consider the effect of introducing a small surface patch at some distance \emph{above} an existing visible surface.
This modification also impacts the rendered image and can be interpreted as a perturbation of a more general \emph{non-local} derivative.
A similar concept was previously used by \citet{mehta2023theory} to \emph{nucleate} new shapes in 2D vector graphics, and by \citet{zhang2024many} in the context of physically based rendering.

Optimizing surfaces on this extended domain mitigates two key issues discussed previously:
Because updates are no longer constrained to the surface, the algorithm can achieve faster and more robust convergence within a higher-dimensional loss landscape, as illustrated below:
\begin{center}
    \vspace{2mm}
    \includegraphics[width=0.96\columnwidth]{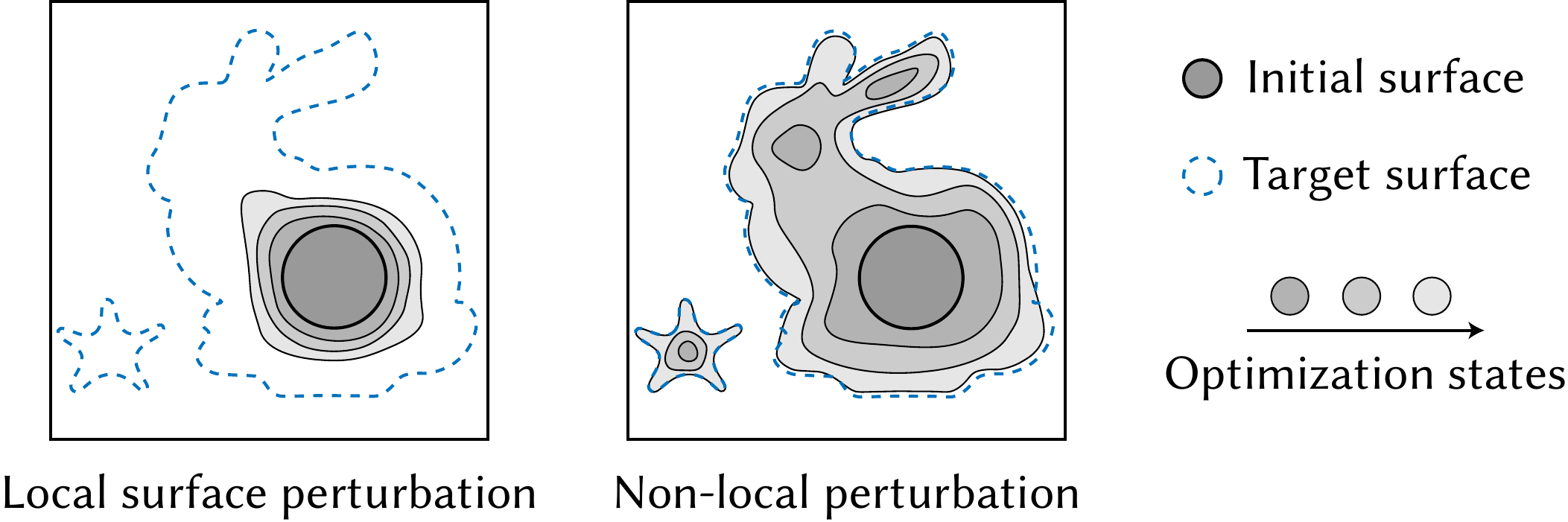}
    \vspace{0mm}
\end{center}
Second, the need for complex, specialized methods to estimate boundary derivatives is eliminated, which simplifies
the implementation and further improves performance.
Before making these abstract notions concrete, we cover the \mbox{used geometric representation.}

\paragraph{Geometric representation}
Non-local perturbations require a representation that spans the entire space.
To this end, we use an \emph{occupancy field} \citep{mescheder2019occupancy,niemeyer2020differentiable} that encodes the discrete probability of a position $\vx$ being occupied:
\begin{align*}
    \occupancy(\vx) = \Pr\{\vx \text{ lies within an object}\} \in [0, 1].
\end{align*}
After convergence, the field is expected to have occupancy values approaching $1$ on the surface, and $0$ in the exterior.
We note that the choice of an occupancy field is somewhat arbitrary. The primary focus of this work is on optimizing geometry irrespective of the specific details of the representation.


\subsection{Radiance field loss}
\label{sec:radiance_field_loss}

\begin{figure}[t]
    \centering
    \includegraphics[width=\linewidth]{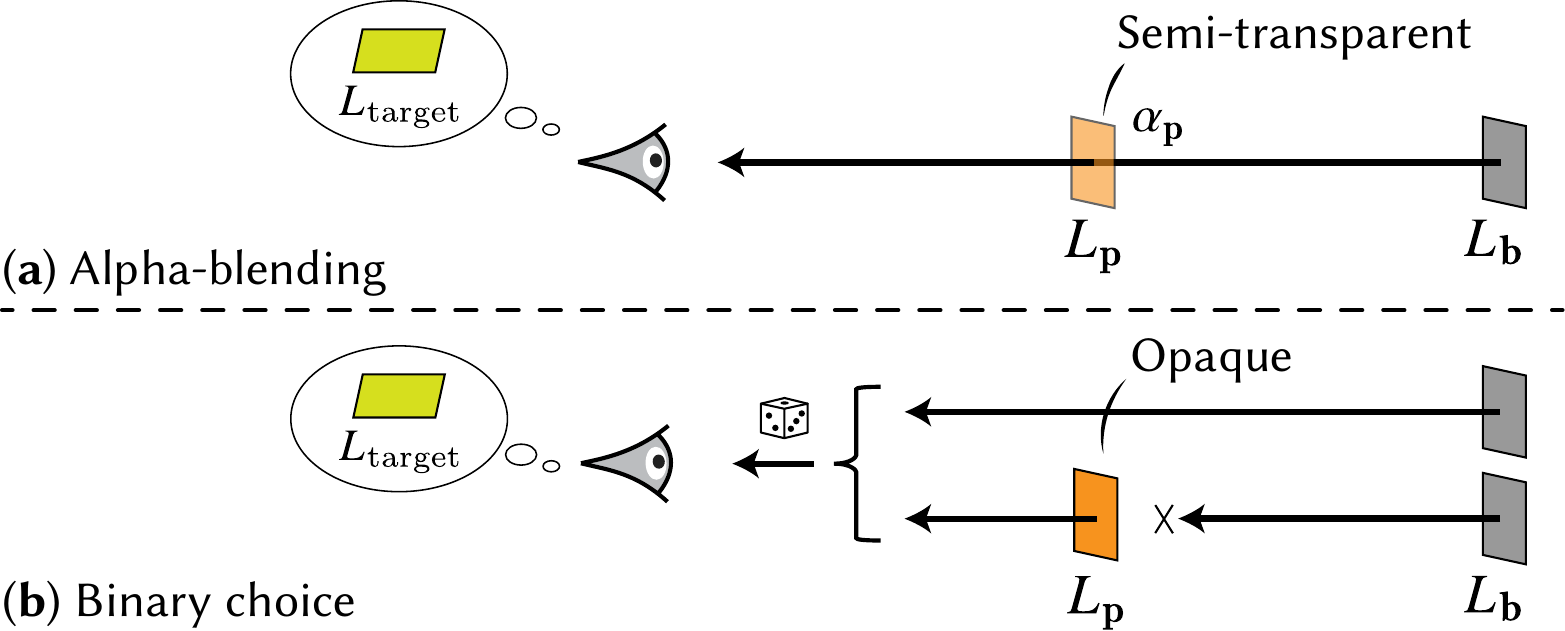}
    \caption{\label{fig:single_candidate}%
        \textbf{Non-local perturbations.}
        We consider a single candidate surface patch (with color $L_\p$) along the ray as a perturbation of a background surface (with color $\Lb$).
        \textbf{(a)} Blending colors violates the surface assumption and leads to volumetric results.
        \textbf{(b)} We instead treat the perturbation as a random binary choice and optimize the associated discrete probability. The final reconstruction is non-random and will never blend the contribution of multiple surfaces.
    }
\end{figure}

\paragraph{Single candidate}
To explain the concept of a non-local perturbation, we first focus on the case of a single candidate surface patch along a ray.
\autoref{fig:single_candidate} depicts this setup, in which a candidate at position $\p$ with color $\Lp$ and occupancy $\occupancy_\p$ precedes a
background\footnote{For now, the term \emph{background} could refer to a surface, an environment map, etc. Later sections will provide a concrete definition.} \mbox{with color $\Lb$.}
How this geometric configuration arises will be covered later---for now, we assume that is given, and that
the color values $\Lp$ and $\Lb$ are furthermore fixed.

In this case, the optimal reconstruction is straightforward: the candidate should be created if it improves the match with respect to a specified target color $L_\mathrm{target}$; otherwise, it should be discarded.

The occupancy parameter $\occupancy_\p$ provides the means to achieve this outcome.
However, there are different ways to integrate it.
The standard volumetric approach (\autoref{fig:single_candidate}a) interprets $\occupancy_\p$ as an \emph{opacity} for alpha-compositing, minimizing a color difference \mbox{$\ell(\hat{L}, L)$ of the form}
\begin{align}
    \ell \left( \occupancy_\p\, \Lp + (1 - \occupancy_\p)\, \Lb,
        ~L_\mathrm{target} \right).
\end{align}
The fundamental limitation of this approach is its inability to promote binary occupancy values.
When the best match is given by a blend of $\Lp$ and $\Lb$, the loss will reach zero without forming a distinct surface.
A common remedy involves adding additional loss terms to penalize such behavior, but this
lacks a principled theoretical foundation and adds complexity in the form of hyperparameters.

We instead interpret the non-local perturbation as a binary choice: the candidate surface either exists, or it does not.
Thus, the final color value associated with the ray is either
that of the candidate $L_\p$ or the background $\Lb$ (\autoref{fig:single_candidate}b).
We quantify the quality of each possibility via $\ell$ and seek the \mbox{occupancy value $\occupancy_\p\in[0,1]$ that minimizes:}
\begin{align}
    \label{eq:subproblem_loss}
    \loss(\p) = \occupancy_\p\, \ell(\Lp, ~ L_\mathrm{target}) +
        (1 - \occupancy_\p)\, \ell(\Lb, ~ L_\mathrm{target}).
\end{align}
By blending the losses of the two surfaces instead of their colors, this
approach selects the surface that best explains the target color.

The simplified example shown here assumes that the candidate color $\Lp$ is static.
In practice, $\Lp$ (but not $\Lb$) is also subject to optimization, which requires multiple viewpoints to resolve ambiguity; more on this later.

\paragraph{Multiple candidates}
We now extend the loss formulation to consider multiple candidates. This is
advantageous because it will allow our method to simultaneously evaluate the effect of several
perturbations, which in turn accelerates convergence.

\begin{figure}[t]
    \centering
    \includegraphics[width=0.85\linewidth]{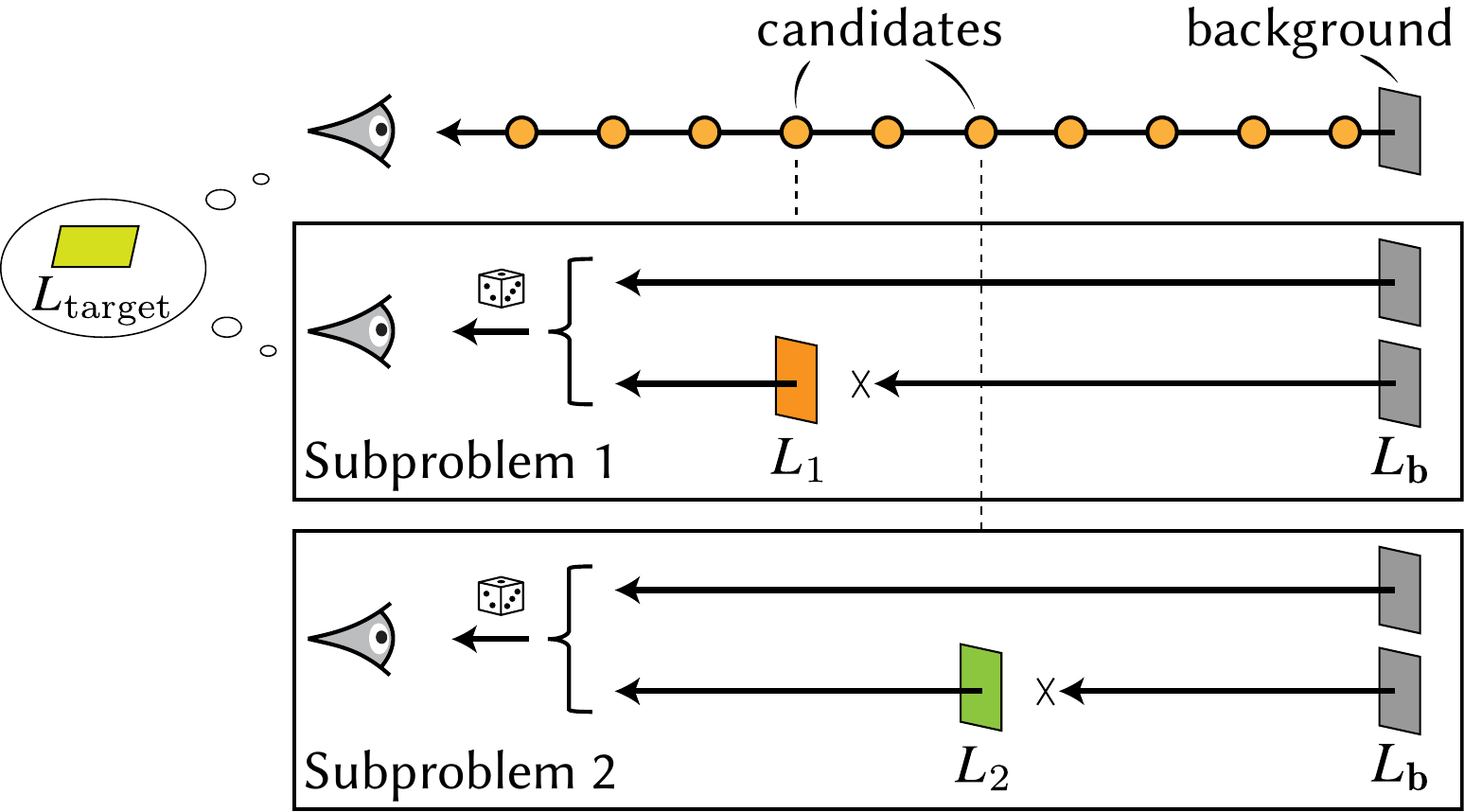}
    \caption{\label{fig:multiple_candidates}%
        \textbf{Surface candidates as independent subproblems.}
        With multiple candidates along a ray, each perturbation is treated as an independent subproblem, resulting in \emph{local} losses distributed spatially over the scene.
    }
\end{figure}

The key property of the single-candidate loss formulation is that it isolates
the candidate from the background surface (i.e., observing one or the other). The
generalization to multiple candidates preserves this property by treating each
candidate as an independent subproblem (\autoref{fig:multiple_candidates}),
minimizing the sum of respective losses:
\begin{align}
    \label{eq:sum_of_subproblems}
    \loss_\mathrm{ray}(\mathbf{r}) =
        \sum_{i=1}^m \loss(\p_i),
\end{align}
where $\loss(\p_i)$
(following Equation~\ref{eq:subproblem_loss})
represents the loss of the $i$-th of $m$ candidates sampled along the ray $\mathbf{r}$.

\paragraph{Spatio-directional loss}
Reconstruction tasks evaluate the loss~\eqref{eq:sum_of_subproblems} along a large set of rays $\mathbf{r}_k\ (k=1, \ldots, n)$, where $n$ denotes the total number of pixels across all reference images. This further expands the set of independently considered candidate surfaces and leads to the combined loss
\begin{align}
    \label{eq:overall_loss}
    \loss_\mathrm{total} = \sum_{k=1}^n\loss_{\text{ray}}(\mathbf{r}_k).
\end{align}

Whereas conventional surface optimization only propagates gradients to the
surface itself, the use of $\alpha_\p$ and $L_\p$ in Equation~\eqref{eq:subproblem_loss} covers the
\emph{entirety} of the observed 3D space. For positions viewed from multiple
directions, the loss generally also varies with respect to direction:
\begin{center}
    \includegraphics[width=0.6\columnwidth]{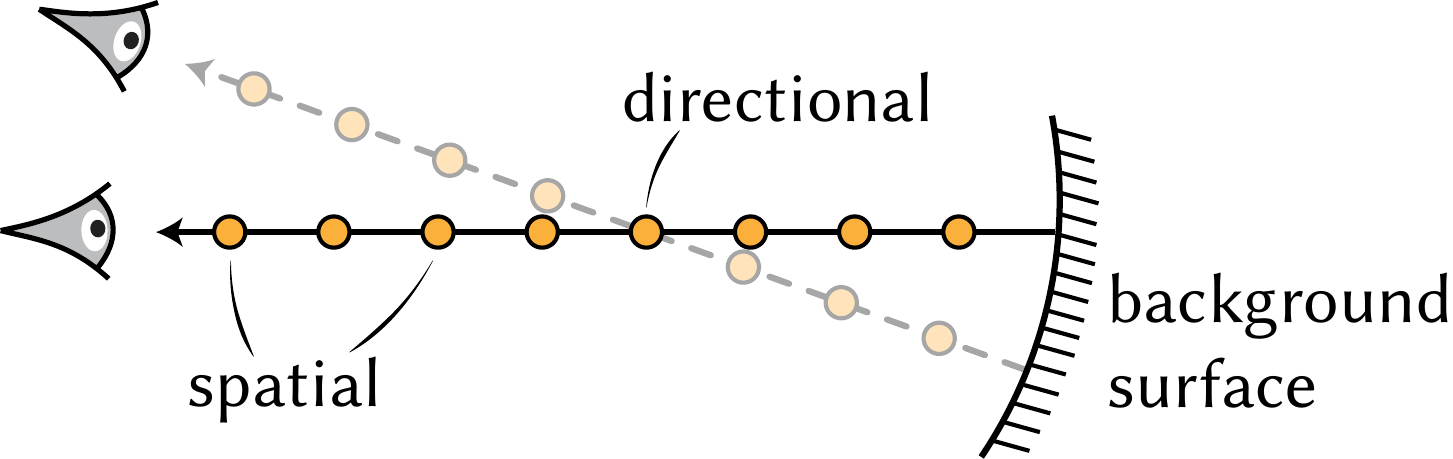}
\end{center}
In other words: by moving the evaluation of $\ell$ from image space into the scene, we have created a
\mbox{spatio-directional \emph{radiance field loss}.}


\subsection{Stochastic background surface}
\label{sec:stochastic_baseline}

\begin{figure}[t]
    \centering
    \includegraphics[width=\linewidth]{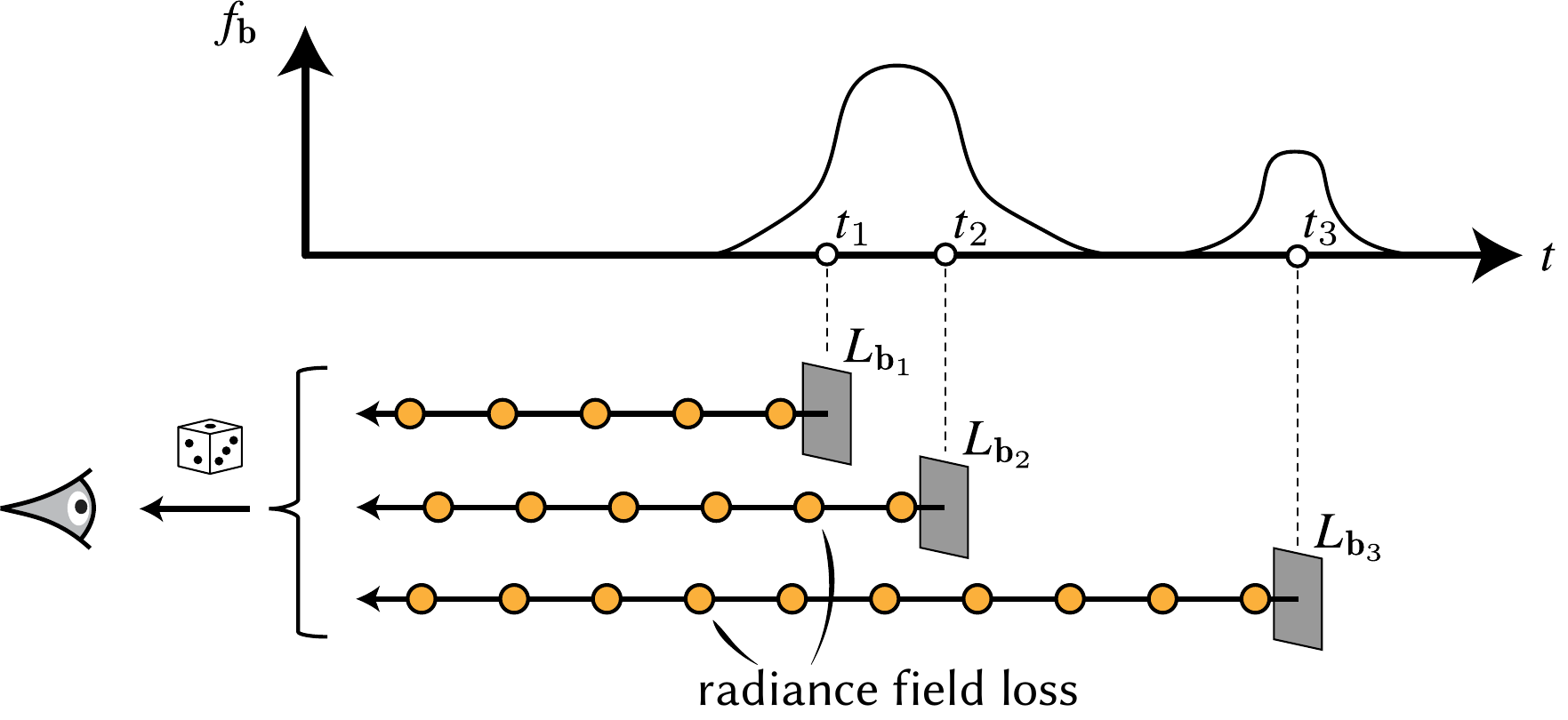}
    \caption{\label{fig:sto_bg}
        \textbf{Stochastic background.}
        Selecting the background surface at random from a distribution $\fb$ enables visibility through high-occupancy regions.
        Each sampled background surface defines a new perturbation problem solvable with the radiance field loss.
        Taking an expectation of this process leads to a simple deterministic expression that we implement in practice.
    }
    \vspace{-4mm}
\end{figure}

To complete our derivation of the loss function, what remains is the definition of the background surface.
Rather than a deterministic surface (e.g., a level set of the occupancy field), we draw the background from a per-ray distribution $\fb$.
This enables occasional ``visibility'' through high-occupancy regions, allowing occluded objects to be considered as the background (\autoref{fig:sto_bg}).
Crucially, we thereby support complex topological changes in our optimization without having to explicitly account for them~\citep{mehta2023theory}; see \autoref{sec:additional_experiments_and_results} for additional details.

The design of the distribution $\fb$ is flexible.
One straightforward approach is to prioritize sampling in high-occupancy regions, as these areas are more likely to correspond to surfaces.
During ray traversal, we stochastically decide whether to use a position as the background surface based on its occupancy value.
This sequential decision process reflects the concept of free-flight distance~\cite{novak18monte}, forming the \emph{free-flight background distribution}.

We can formulate the expectation of sampling the background surface from such a free-flight distribution analytically and derive a corresponding aggregated local loss analogous to classical volumetric light transport:
\begin{align}
    \label{eq:expectation_loss_ray_march_simp}
    \loss(\p_i)
    &=
        \left( \prod_{j=1}^{i - 1} (1 - \alpha_{\p_j}) \right)
         \alpha_{\p_i} \ell(L_{\p_i})  \,,
\end{align}
which, when plugged into Equation~\eqref{eq:overall_loss}, yields the radiance field loss~(\autoref{fig:nerf_comparison}).
See \autoref{sec:expectation_over_bg} for the complete derivation.

\paragraph{Implementation}
An implementation of our loss function can be arranged to resemble the color blending structure of standard volume reconstruction methods like NeRF~\cite{mildenhall2020nerf}.
As such, it is exceedingly simple to implement in existing codebases, as illustrated in the following comparison of pseudocode.
\begin{center}
    \includegraphics[width=0.9\columnwidth]{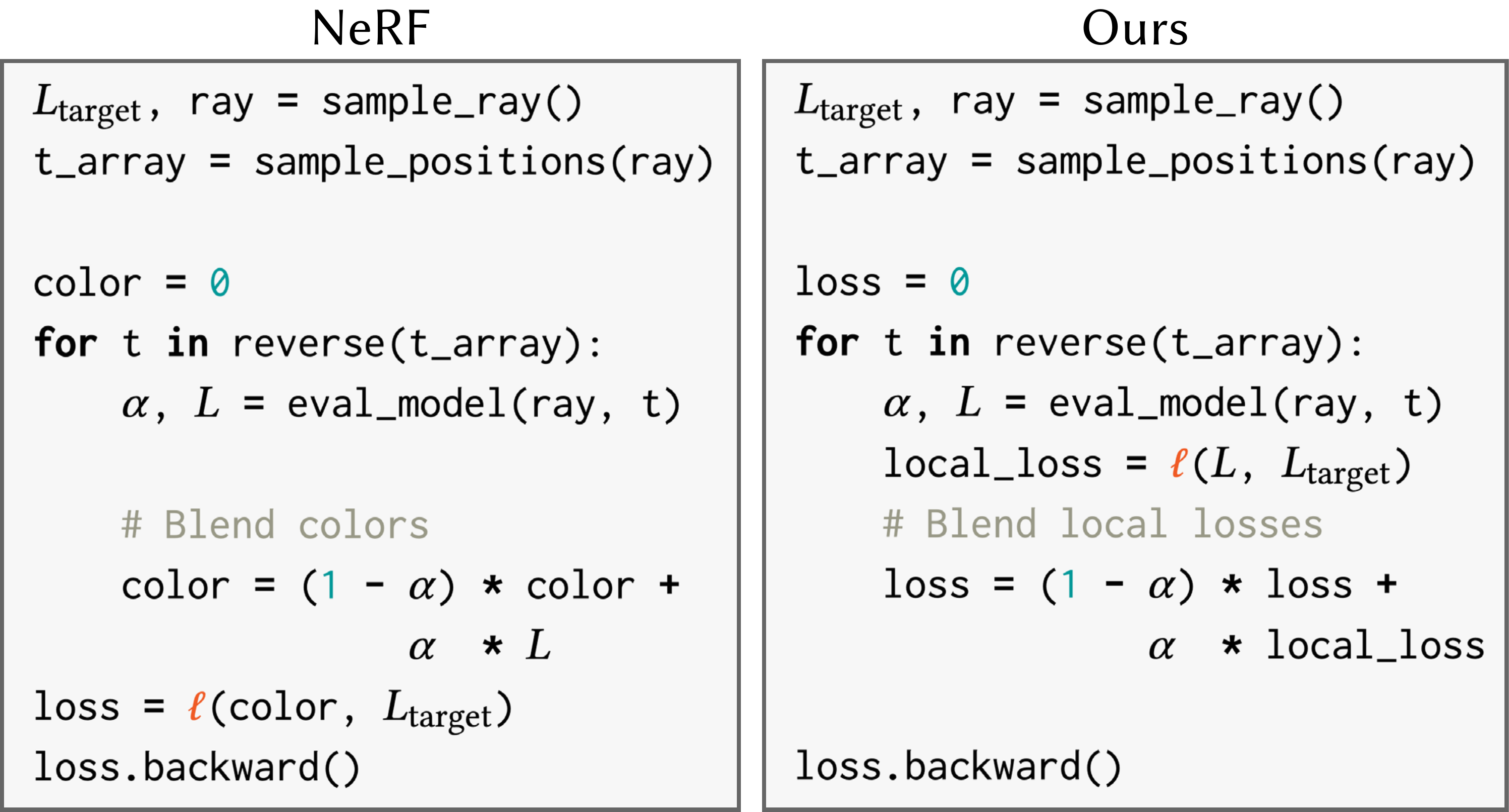}
\end{center}
This resemblance also suggests that the optimization landscape of our method is similar to that of NeRF, inheriting its robustness.
However, while NeRF's loss supervises all samples along the ray to collectively match the target color, our loss aims for each sample to match the target color independently or become transparent when the background is a better match.
This distinction fundamentally defines our approach as a surface reconstruction algorithm.

\subsection{Volume relaxation}
\label{sec:relaxed_variant}

We also propose a heuristic-based generalization of our method.
It is orthogonal to the above algorithm and optional during training.

While a surface representation offers many advantages, the opaque surface assumption has inherent limitations in certain scenarios.
For example, sub-pixel structures are challenging to model with geometry, and a single surface may fail to accurately represent the appearance of directional-varying materials.
In these regions, a volumetric representation is more suitable.

Our goal is to relax our method to reconstruct most of the scene as surfaces (regions where low loss can be reached) and use volumetric representations only in the remaining challenging regions.
To this end, we first train with our algorithm for $20$k iterations to obtain an initial surface representation.
We then identify challenging regions by evaluating where local losses remain high.
In subsequent training steps, we relax the surface assumption, allowing volumetric alpha blending in these regions.

After training, rather than extracting a surface, we render the scene volumetrically, with surface regions treated as fully opaque ``volumes''.
Comparing to a volume scene optimized with NeRF, our method still benefits from the compact representation of surface regions.
When accumulating colors along a ray, very few samples are required to saturate the transmittance, leading to faster inference and reduced computational resources during training.

Where not explicitly stated, all results in this paper (marked as ``ours'') are trained \emph{without} volume relaxation.


\section{Results}
\label{sec:results}

\subsection{Novel view synthesis}

\begin{figure*}[ht]
    \centering
    \includegraphics[width=\textwidth]{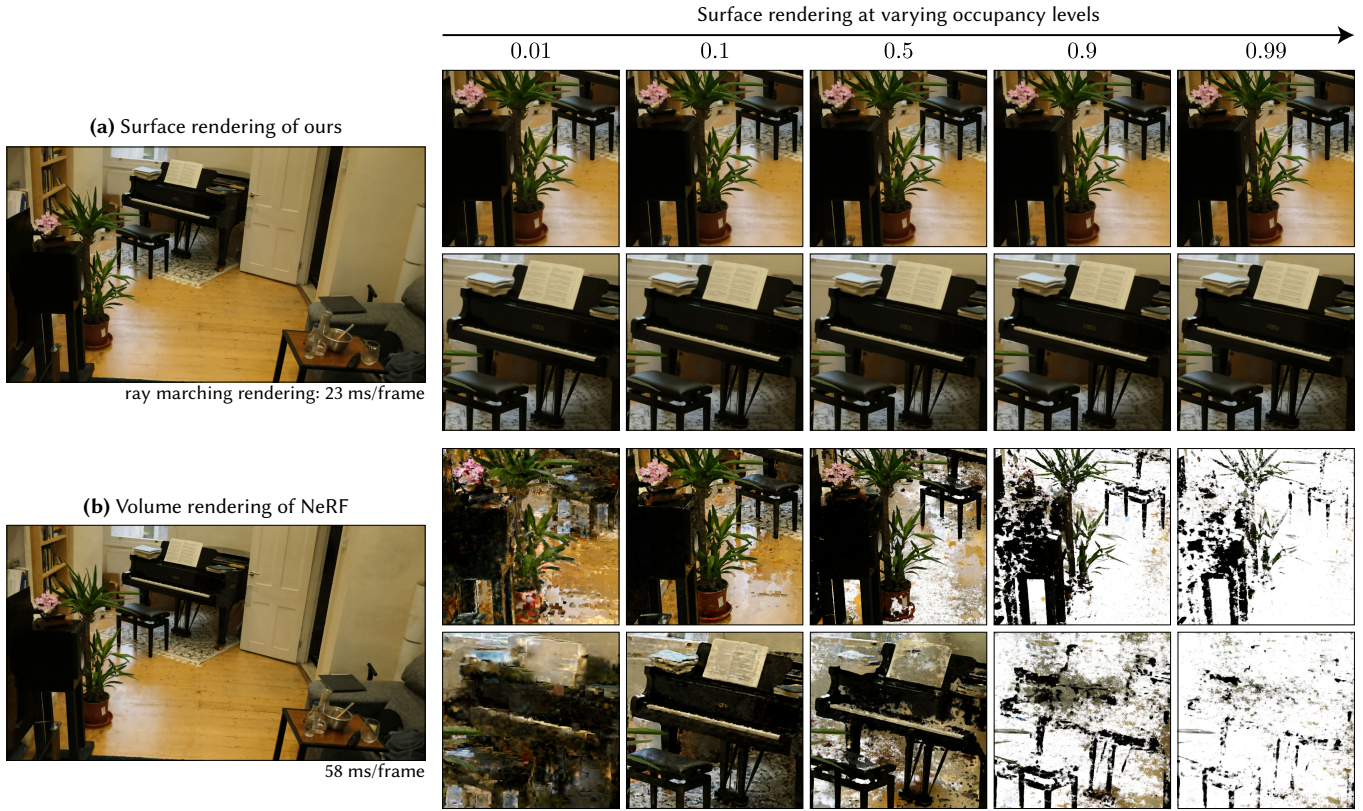}
    \caption{
        \textbf{Nature of the reconstructed occupancy.}
        Surface rendering at varying level sets of a scene reconstructed by our method and NeRF, both implemented in Instant NGP using the same hyperparameters.
        \textbf{(a)} For our method, the surface rendering shows minimal changes across different level set thresholds, indicating that the occupancy field has converged to a near-Heaviside step function on the surface, allowing for extraction of a surface-based representation.
        \textbf{(b)} NeRF reconstructs the scene volumetrically, and any surface extracted using a level set is a poor approximation of the true color.
    }\label{fig:level_sets_room}
\end{figure*}

\begin{figure*}[ht]
    \centering
    \includegraphics[width=\textwidth]{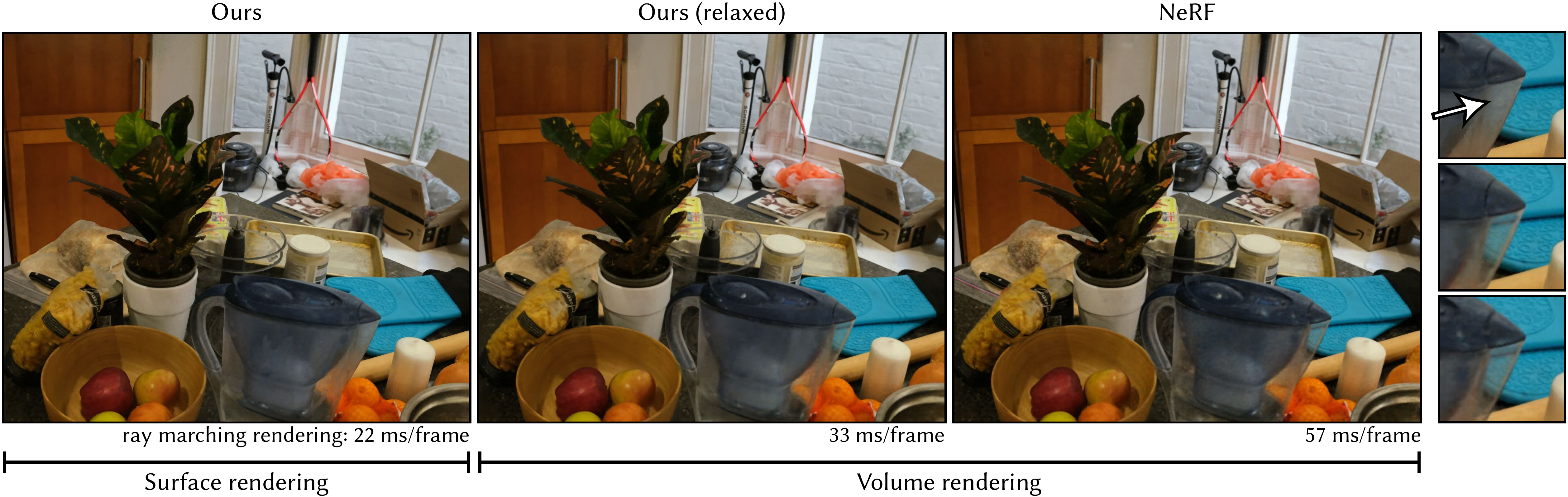}
    \caption{
        \textbf{Volumetric relaxation.}
        We compare reconstructions of our method \emph{without} and \emph{with} volume relaxation to NeRF, all implemented in Instant NGP using the same hyperparameters.
        While our method achieves comparable visual quality using a surface-based representation, we highlight a region (white arrow) where it fails to model a semi-transparent object due to the opaque surface assumption.
        The relaxed variant of our algorithm can recover by adopting volume rendering in such regions.
        Rendering the reconstructions using the same ray marching implementation leads to significant performance differences:
        our surface-only reconstruction is $2.6\times$ faster than NeRF\@.
        The relaxed variant benefits from the surface representation in most regions, and is $1.7\times$ faster.
    }\label{fig:relaxed}
\end{figure*}

\paragraph{Visual quality}
Despite the inherently fewer degrees of freedom of surfaces, \autoref{fig:level_sets_room} shows that our method achieves results that are qualitatively comparable to NeRF\@.
We also visualize the surface renderings at occupancy level sets $\{0.01, 0.1, 0.5, 0.9, 0.99\}$.
Renderings of the scene optimized by our algorithm barely change, indicating a near-Heaviside step function in the occupancy field.
In contrast, the inherently volumetric nature of NeRF does not produce meaningful visualizations for these level sets.
\autoref{fig:relaxed} highlights the reconstruction of another scene where our method with volume relaxation addresses challenges in modeling a \mbox{semi-transparent object.}

\autoref{tab:metric_mean} shows that our method achieves visual quality comparable to exponential volume reconstruction (NeRF) when trained on the MipNeRF360 dataset, using default Instant NGP hyperparameters, despite using a surface-based representation.
A small PSNR gap is expected, as volume representations offer inherently more degrees of freedom that can be repurposed to model pixel-wise colors.
A similar trend is observed when implementing our method in the ZipNeRF codebase, where we measured mean PSNR of $29.73$ dB for our method and $31.45$ dB for NeRF on indoor scenes, and $24.06$ dB and $25.24$ dB on outdoor scenes, respectively.

When evaluating our relaxed variant---which switches to volumetric rendering in hard regions---the visual quality slightly exceeds the NeRF baseline.
This improvement arises because our method encourages surface-like, sparse distributions, resulting in more empty space that the renderer can efficiently skip.
Consequently, at equal batch size, Instant NGP automatically spawns more rays when using our method, thereby covering more reference pixels per batch, in turn leading to a better reconstruction.
When the ray count is restricted to match NeRF, the relaxed variant delivers results that are approximately equal.

These trends are consistent across other metrics as well. For instance, both our method and NeRF achieve SSIM scores of $0.89$ (indoor) and $0.68$ (outdoor). The full set of evaluation results is provided in the appendix.

\paragraph{Rendering performance}
Our implementation builds on the Instant NGP codebase, which ray-marches fields ($\alpha_\p, L_\p$) represented using an interpolated hash grid lookup combined with a lightweight MLP.
We repurpose this ray-marching code for surface rendering by returning the color of the first sample with an occupancy value exceeding $0.5$.
This straightforward modification results in a $2.4\times$ average speedup in frames per second (FPS) across MipNeRF360 scenes compared to the baseline.
An average speedup of $2.0\times$ is achieved for the relaxed version of our method, as most of the scene remains surface-like.

An additional $2\times$ speedup can be achieved by replacing ray-marching with rasterization of a meshed isosurface.
In this case, the color network is only used for mesh shading, maintaining the same visual quality as before.
Various strategies exist to further boost rendering performance, e.g., by storing precomputed hash grid lookups alongside mesh vertices~\cite{chen2023mobilenerf}, or by projecting the directional MLP dependence into \mbox{spherical harmonics~\citep{reiser2024binary}.}

\begin{table}[t]
    \caption{\textbf{Visual quality comparison}. We integrate our loss into Instant NGP and train on the MipNeRF360 dataset using default hyperparams.}
    \label{tab:metric_mean}
    \begin{tabular}{@{\hskip -0.7mm}l@{\hskip 2.5mm}c@{\hskip 2.5mm}c@{\hskip 2.5mm}c@{\hskip 2.5mm}c@{\hskip 2.5mm}c@{\hskip 2.5mm}c@{\hskip -0.2mm}}
    \toprule
                    & \multicolumn{3}{c}{Indoor mean} & \multicolumn{3}{c}{Outdoor mean} \\
                    \cmidrule(lr){2-4} \cmidrule(lr){5-7}
                    & PSNR$\uparrow$ & SSIM$\uparrow$ & LPIPS$\downarrow$ & PSNR$\uparrow$ & SSIM$\uparrow$ & LPIPS$\downarrow$  \\
    \midrule
    Ours            & 29.02 dB & 0.888 & $\bm{0.275}$ & 22.41 dB & 0.679 & $\bm{0.563}$ \\
    Ours (relaxed)  & $\bm{29.41}\,\mathrm{\mathbf{dB}}$ & $\bm{0.897}$ & 0.284 & $\bm{22.62}\,\mathrm{\mathbf{dB}}$& $\bm{0.690}$ & 0.626 \\
    NeRF            & 29.19 dB & 0.893 & 0.303 & 22.47 dB & 0.683 & 0.638 \\
    \bottomrule
    \end{tabular}
\end{table}

\begin{figure}[t]
    \centering
    \includegraphics[width=1.0\linewidth]{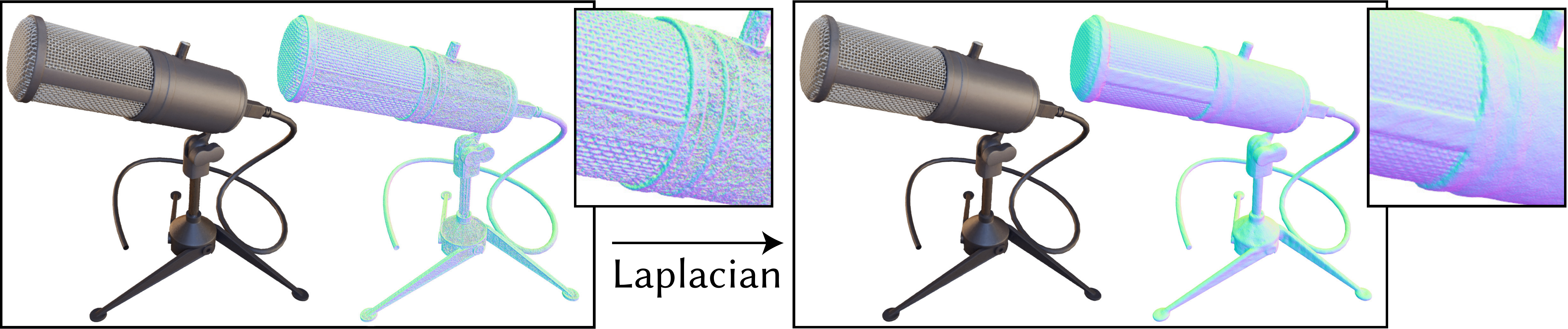}
    \caption{
        \textbf{Regularization}.
        For simple scenes with enough observations, the reconstructed surface closely matches the ground truth geometry without requiring additional constraints.
        Adding Laplacian refinement helps smooth out unnecessary small kinks, resulting in a more accurate final geometry.
    }\label{fig:simple_lap}
\end{figure}

\subsection{Geometry reconstruction}

\begin{figure}[t]
    \centering
    \includegraphics[width=1.0\linewidth]{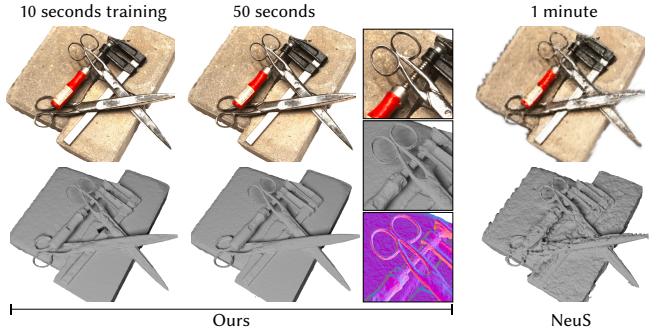}
    \caption{
        \textbf{Straightforward extraction.}
        Since our algorithm does not use an intermediate volume representation, efficient surface extraction is possible at any point.
        At equal time, a fast NeuS2 baseline~\citep{wang2023neus2} still models the scene as a fuzzy volume, and a surface \mbox{cannot be confidently extracted.}
    }\label{fig:neus}
\end{figure}

Our method is also applicable to geometry reconstruction tasks, in which achieving high-quality meshes matching the ground truth geometry is of interest.
For simple multi-view input (\autoref{fig:simple_lap}), our method produces highly detailed geometry with the speed of Instant NGP (seconds). A mesh can be extracted at any point during the optimization (\autoref{fig:neus}).
However, complex real-world reconstruction tasks are often under-constrained.
For example, a reflective object seen only from a narrow cone of directions does not provide sufficient information for accurate shape recovery.
Even with a larger set of viewpoints, it can be challenging to disambiguate whether surface detail is due to local color variation or small-scale geometry.
As a consequence, the reconstructed geometry often exhibits undesirable bump-like artifacts representing such misattributed detail.
While our algorithm still excels at novel view synthesis under these conditions,
the reconstructed geometry can significantly deviate from the ground truth.

\begin{table}[t]
    \caption{Average Chamfer distance comparison on the DTU dataset
    with NeuS \cite{wang2021neus} and NeuS2 \cite{wang2023neus2}.}
    \vspace{-2mm}
    \label{tab:cd_average}
    \centering
    \begin{tabular}{lccc}
    \toprule
       & Ours (1 min) & NeuS (8 hr) & NeuS2 (5 min) \\
    \midrule
    CD & 0.80 & 0.77 & $\bm{0.68}$ \\
    \bottomrule
    \end{tabular}
\end{table}

To mitigate this issue, we incorporate an exponentially decaying Laplacian regularizer during training.
This regularizer initially enforces flat surfaces
and progressively provides more degrees of freedom as its influence decays.
\autoref{fig:laplacian_decay} examines the influence of the final Laplacian weight on reconstruction quality.
\autoref{fig:geometry_big} showcases geometry reconstructions of scenes from the
DTU~\cite{jensen2014large} and BlendedMVS~\cite{yao2020blendedmvs} datasets, all made
with a consistent
Laplacian weight of $2\times 10^{-5}$.

\autoref{tab:cd_average} shows that using only minimal Laplacian regularization, our method achieves an average Chamfer distance on the DTU dataset that is just $0.12$ higher than NeuS2~\cite{wang2023neus2}, while reducing runtime to only 1 minute thanks to our algorithmic simplicity.
The complete evaluation results are provided in the appendix.
In this work, we do not intend to compete on geometric reconstruction metrics and have not incorporated other regularization extensions, which
would detract from the simplicity of the presented idea.
Such extensions include multi-view consistency losses \cite{fu2022geo,chen2024pgsr} to reduce ambiguities in regions with limited observations,
or the TSDF algorithm \cite{izadi2011kinectfusion} that helps extract smooth meshes while removing unnecessary geometry.


\section{Discussion}
\label{sec:discussion}

\subsection{Choice of background distribution}
In \autoref{sec:stochastic_baseline}, we used the free-flight background distribution to derive a loss form dual to the NeRF loss. This choice is somewhat arbitrary, and other distributions could be used with different trade-offs.
In \autoref{sec:choice_of_distribution}, we discuss how alternative designs can enable new optimization strategies that are not possible in image-space methods, with one such example provided.

\subsection{Relation to many-worlds inverse rendering}
\label{sec:relation_to_many_worlds}

Our method builds on the core idea of~\citet{zhang2024many} (we refer to their
method as \emph{PBR-MW}), namely that surface distributions can be optimized
more directly without involving exponential volumes. We reconstruct purely
emissive objects\footnote{In the equations of physically based rendering, radiance fields manifest in the emission term~\cite{nimierdavid2022unbiased}.}, while PBR-MW handles differentiable shadowing and
interreflection to reconstruct reflecting objects in scenes with global
illumination. Viewed superficially, our method could be mistaken for a stripped
down version of PBR-MW\@.

Our contribution lies in leveraging this simplicity to develop a specialized
method. We identify and implement optimizations unique to radiance surfaces to
fully realize the potential of the many-worlds idea.

In radiance surface rendering, image formation is a direct 1:1 mapping
between a ray and the nearest intersected surface, while PBR-MW
requires a complex nested integration over materials, lighting, and
geometry. Our approach to project training images into the scene to establish a
radiance field loss depends on this 1:1 mapping and does not efficiently
translate to the nested integral structure of a global illumination renderer.

Another important contribution is the introduction of a stochastic background
distribution, which enables topological changes and substantially improves
reconstruction quality. We show how to cheaply evaluate this strategy in
expectation, which is needed to maintain algorithmic parity with NeRF\@. The
associated derivations and simplifications (\autoref{sec:expectation_over_bg})
are specific to radiance surfaces and do not transfer to PBR-MW\@.

\subsection{Limitations and future work}
Moving the evaluation of the color loss $\ell$ from image space into the radiance field makes our method incompatible with loss functions that depend on image-space neighborhoods (e.g., style losses).

\begin{figure}[t]
    \centering
    \includegraphics[width=0.98\linewidth]{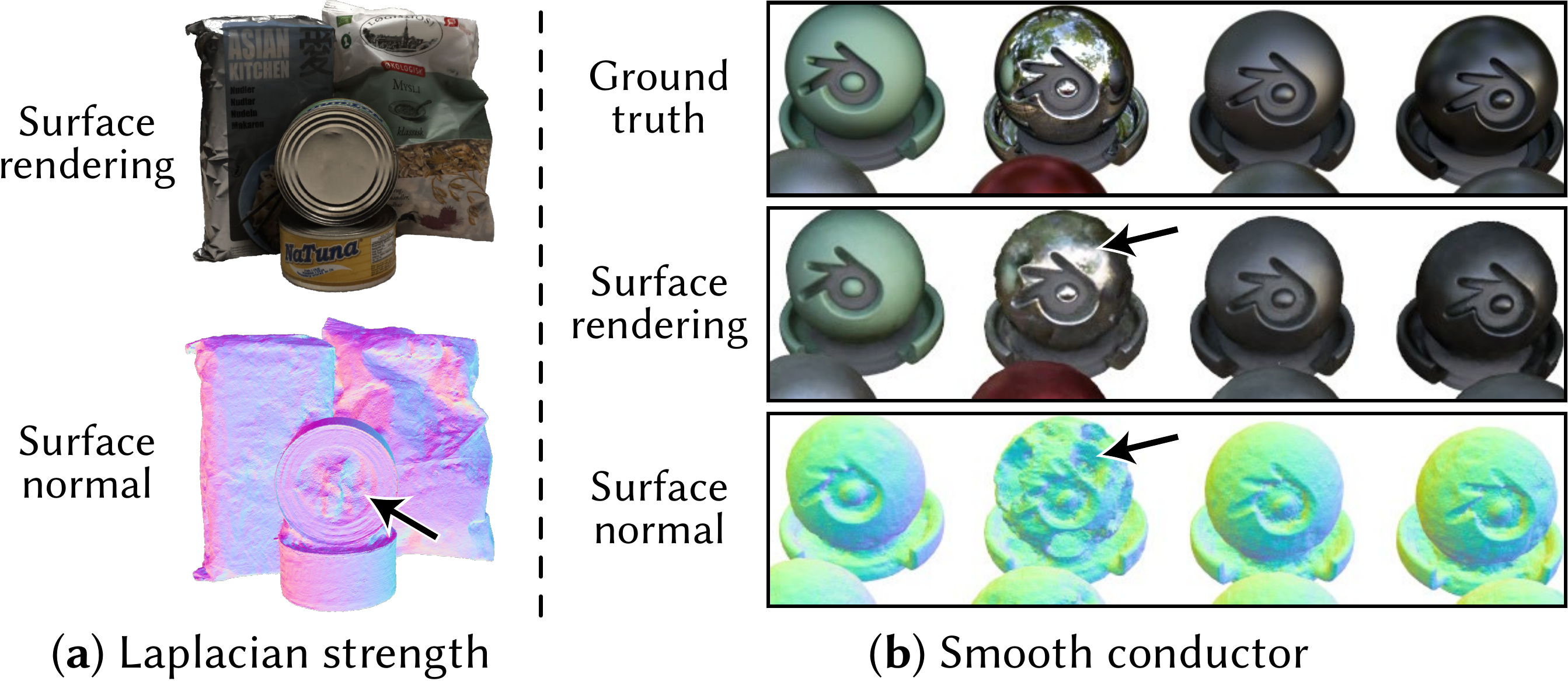}
    \caption{
        \textbf{Limitations. (a)} Our Laplacian smoothing strategy fails to reconstruct the flat can surface due to its view-dependent appearance.
        A larger Laplacian weight can help, but this also
        suppresses geometric detail seen in \autoref{fig:geometry_big}.
        \textbf{(b)} High-frequency color variation is more challenging to accurately represent on a surface compared to a volumetric representation.
    }\label{fig:limitation}
\end{figure}

As shown in \autoref{fig:limitation}, our lightweight Laplacian regularization fails when there are insufficient observations to constrain the geometry.
Using alternative regularization techniques from state-of-the-art geometry reconstruction methods could help mitigate this issue.
Our method also struggles to accurately capture the appearance of conductive materials, which could be addressed by incorporating solutions from prior work \citep{verbin2021refnerf}.

An interesting extension of our work could involve implementing a particle-based storage approach, such as Gaussian splatting \citep{kerbl20233d}.
However, 3D Gaussians are inherently semi-transparent, which conflicts with our assumption of opacity. Future work could explore the use of opaque primitives, such as 2D disks, to replace semi-transparent particles.

\section{Conclusion}
\label{sec:conclusion}

The "many worlds" paradigm---i.e., optimizing a distribution over non-interacting primitives---is relatively new in the field of differentiable rendering.
In this paper, we apply it to radiance surface reconstruction, which yields a fast and simple alternative to prior works.
Particularly notable is that the derivation began with an evolving surface, yet resulted in remarkably similar equations to volumetric scene reconstructions: ones where losses rather than colors are integrated along rays.

As reconstruction tasks increase in difficulty, a key challenge lies in deciding whether a region of space is best represented by a surface or a volume.
While the relaxed variant of our method offers an effective heuristic, it also underscores the need for a principled answer to this important question.

Much engineering has gone into the design of optimized algorithms, regularizers, and heuristics for NeRF-based 3D reconstruction.
Our hope is that a large portion of this effort will translate to the radiance field loss and yield state-of-the-art results in the future.

\begin{acks}
    The authors would like to thank Aaron Lefohn and Alexander Keller for their support.
    This project has received funding from the European Research Council (ERC) under the European Union's Horizon 2020 research and innovation program (grant agreement No 948846).
\end{acks}


\bibliographystyle{ACM-Reference-Format}
\bibliography{main.bib}


\begin{figure*}[ht]
    \centering
    \includegraphics[width=0.95\textwidth]{images/laplacian_decay/laplacian_decay.pdf}
    \vspace{-2mm}
    \caption{
        \textbf{Effect of Laplacian weight on geometry reconstruction.} The above results demonstrate the trade-off between geometric detail and surface smoothness. For simple scenes lacking intricate features (bottom row), the reconstruction is insensitive to this hyperparameter.
    }
    \label{fig:laplacian_decay}
\end{figure*}
\begin{figure*}[ht]
    \centering
    \includegraphics[width=0.89\textwidth]{images/geometry_reconstruction/geometry.pdf}
    \vspace{-2mm}
    \caption{
        \textbf{Reconstruction showcase.}
        Surface rendering and normals of various scenes from the DTU and BlendedMVS datasets,
        reconstructed with our algorithm and a decaying Laplacian.
        All results were generated with the same hyperparameters and a training time of $\sim 1$ minute per scene.
    }\label{fig:geometry_big}
\end{figure*}


\clearpage

\appendix

\section{Derivation of our loss in NeRF form}
\label{sec:expectation_over_bg}

\begin{figure}[h]
    \centering
    \includegraphics[width=1.0\linewidth]{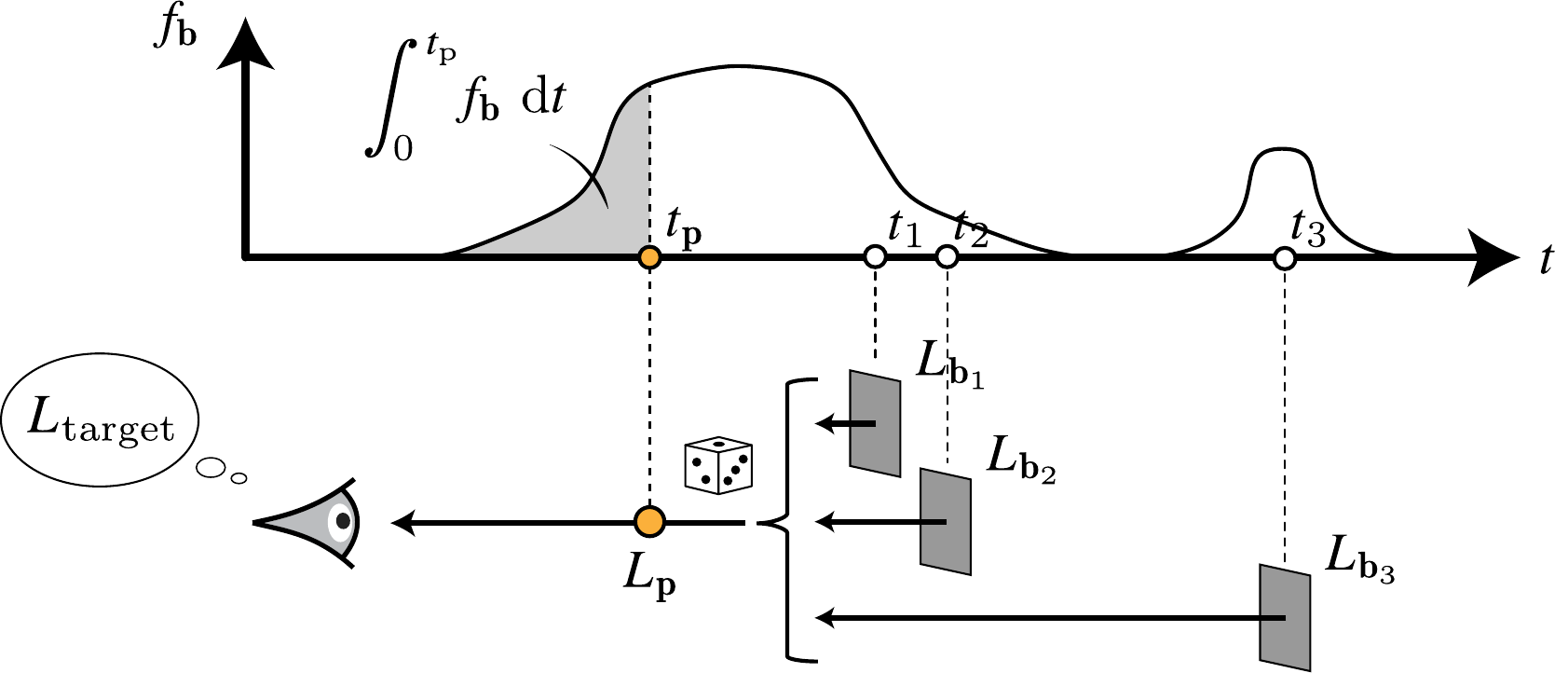}
    \caption{
        We derive an analytic loss expectation over all possible background surfaces for a specific candidate position at $t_\p$.
    }
    \label{fig:appendix_bg}
\end{figure}

Following the introduction of the radiance field loss and the stochastic background, a naive implementation of our method would first sample a surface $\bgsurf$ from a distribution as the perturbation background, and for each sampled $\bgsurf$, we need to sample multiple candidates to solve the non-local perturbation problem.
Naively applying this strategy would result in inefficient implementation.
In the following, we derive an expectation of losses over all potential background surfaces (\autoref{fig:appendix_bg}) for a specific candidate.

Let $\fb$ be the probability distribution of the background surface along a ray, where $\int_{0}^{\tmax} \fb(t) \dt = 1$.
Without loss of generality, we focus on one candidate position $\p$ at distance $t_\p$ along the ray.
To avoid clutter, we denote the color error metric $\ell(L, ~L_\mathrm{target})$ as $\ell(L)$.
The expectation of all losses in the form of \autoref{eq:subproblem_loss} of the main text, local at $\p$, is given by:
\begin{align}
    \E[\loss(\p)]
    &= \int_{t_\p}^{\tmax} \loss(\p) \, \fb(t) \, \dt
    \notag \\
    &=
    \int_{t_\p}^{\tmax} \left( \alpha_\p \, \ell(L_\p) + (1 - \alpha_\p) \, \ell(L_t) \right) \, \fb(t)  \dt
    \notag \\
    &= \left(1 - \int_{0}^{t_\p} \fb(t) \dt\right) \alpha_\p \, \ell(L_\p)  ~ +
        \notag \\
        &\qquad\qquad\qquad\qquad (1 - \alpha_\p)
        \int_{t_\p}^{\tmax} \ell(L_t) \, \fb(t) \dt.
\end{align}
Let
$\E_{t>t_\p} [\ell(L_t)] $ be the expectation of error metrics for $t > t_\p$.
We can rewrite the result as:
\begin{align}
    \label{eq:expectation_loss}
    \E[\loss(\p)]
    &= \Big( \underbrace{1 - \int_{0}^{t_\p} \fb(t) \dt}_{\mathrm{weight}} \Big)
    \Big( \alpha_\p \, \underbrace{\ell(L_\p)}_{\mathrm{candidate}}  +~  (1 - \alpha_\p) \underbrace{\E_{t>t_\p} [\ell(L_t)]}_\mathrm{background} \Big).
\end{align}

\autoref{eq:expectation_loss} reflects an aggregated form of non-local perturbation, where the background color is treated as an expectation over all possible background surfaces rather than a fixed value.
The weight term captures the probability of selecting a background surface located behind the perturbation position. The expectation computation should not change the loss landscape, so the weight term should not be differentiated during optimization.

In the following, we analyze the discrete case of the loss expectation along a ray with $m$ sampled points, assuming the free-flight background distribution $\fb$.
For simplicity, we denote the occupancy and color at position $\p_i$ as $\alpha_i$ and $L_i$, respectively.
The summation of all losses is given by:
\begin{align}
    \label{eq:loss_ray}
    \loss_\mathrm{ray}
    &= \sum_{i=1}^{m} \hat{\E}[\loss(\p_i)]
     \\  \notag
    &= \sum_{i=1}^{m}
        \Bigg[
            \Big(  \underbrace{ \prod_{k=1}^{i - 1} (1 - \alpha_k) }_{\mathrm{weight}}  \Big)
            \Big( \alpha_i \,
            \underbrace{\ell(L_i)}_\mathrm{candidate}
            + (1 - \alpha_i) \,
            \underbrace{\hat{\E}_{t>t_i}[\ell(L_t)]}_\mathrm{background}
             \Big)
        \Bigg],
\end{align}
\vspace*{-4mm}
where
\begin{align}
\label{eq:background_exp}
\hat{\E}_{t>t_i}[\ell(L_t)] =
\sum_{j=i+1}^{m} \left(\prod_{t=i+1}^{j-1} (1 - \alpha_t) \right) \, \alpha_j \, \ell(L_j).
\end{align}
Not all variables in this loss function are meant to be differentiated.
Specifically, the weight term and the background are treated as constants in the optimization process and are excluded from differentiation (i.e., detached).
To indicate which terms should be differentiated, we underline them in the derivation as $\attul{(\cdot)}$.

Below, we reformulate the loss function into a structure similar to the NeRF loss function.
We take some notational liberty of using $\argmin$ to
transform the equation in such a way that its
derivatives and the location of minima are preserved, but the loss value may differ by a constant value.

\autoref{eq:loss_ray} then becomes equivalent to minimizing:
\begin{align}
    \label{eq:loss_ray2}
    &\argmin
    \sum_{i=1}^{m}
    \left[
        \left(
            \prod_{k=1}^{i - 1} \detul{(1 - \alpha_k)}\,
        \right)
        \left(
            \attul{\alpha_i} \, \ell(\attul{L_i}) +
            (1-\attul{\alpha_i}) \,
            \hat{\E}_{t>t_i}[\ell(L_t)]
        \right)
    \right]
    \notag  \\
    =& \argmin
    \underbrace{
    \sum_{i=1}^{m}
        \left(
            \prod_{k=1}^{i-1} (1-\alpha_k)
        \right)
            \alpha_i \, \ell(\attul{L_i}) \,
    }_{\mathrm{(a)}}
    +
    \underbrace{
    \sum_{i=1}^{m}
        \left(
            \prod_{k=1}^{i-1} (1-\alpha_k)
        \right)
            \attul{\alpha_i} \, \ell(L_i) \,
    }_{\mathrm{(b)}}
    \notag   \\
    &\quad\quad  +
    \underbrace{
    \sum_{i=1}^{m}
        \left(
            \prod_{k=1}^{i - 1}\detul{(1-\alpha_k)}
        \right)
        \left(
            (1 - \attul{\alpha_i}) \,
            \hat{\E}_{t>t_i}[\ell(L_t)]
       \right),
    }_{\mathrm{(c)}}
\end{align}
where the terms $\mathrm{(a)}$ and $\mathrm{(b)}$ arise from an application of the product rule.
Reordering the double summation in term $\mathrm{(c)}$ yields:
\begin{align}
    \label{eq:loss_c1}
    \mathrm{(c)} &
    \overset{\eqref{eq:background_exp}}{=}
    \sum_{i=1}^{m}
        \sum_{j=i+1}^{m}
        \left(
            \prod_{k=1}^{i - 1}\detul{(1-\alpha_k)} \,
        \right)
        \left(
            (1 - \attul{\alpha_i}) \,
                \left(
                   \prod_{t=i+1}^{j-1}\detul{(1-\alpha_t)} \,
                \right)
                \detul{\alpha_j \, \ell(L_j)}
       \right)
    \notag \\
    &=
    \sum_{j=1}^{m} \,\,
        \sum_{i=1}^{j-1}
        \,\,\,
        \left(
            \prod_{k=1}^{i - 1}\detul{(1-\alpha_k)} \,
        \right)
        \left(
            (1 - \attul{\alpha_i}) \,
                \left(
                   \prod_{t=i+1}^{j-1}\detul{(1-\alpha_t)} \,
                \right)
                \detul{\alpha_j \, \ell(L_j)}
       \right).
\end{align}
We rename the indices $i \leftrightarrow j$ and subsequently simplify the expression to:
\begin{align}
    \label{eq:loss_c2}
    (\ref{eq:loss_c1})
    & =
    \sum_{i=1}^{m}
        \sum_{j=1}^{i-1}
        \left(
            \prod_{k=1}^{j - 1}\detul{(1-\alpha_k)} \,
        \right)
        \left(
            (1 - \attul{\alpha_j}) \,
                \left(
                   \prod_{t=j+1}^{i-1}\detul{(1-\alpha_t)} \,
                \right)
                \detul{\alpha_i \, \ell(L_i)}
       \right)
    \notag \\
    &=
    \sum_{i=1}^{m}
        \sum_{j=1}^{i-1}
        \prod_{\substack{k=1 \\ k\neq j}}^{i-1}
        \detul{(1-\alpha_k)} \,
        (1 - \attul{\alpha_j}) \,
        \detul{\alpha_i \, \ell(L_i)}.
\end{align}
Combining the term $\mathrm{(c)}$ in the form of \autoref{eq:loss_c2} with the term $\mathrm{(b)}$, we obtain:
\begin{align}
    \label{eq:loss_c3}
    \mathrm{(b)} \!+\! \mathrm{(c)}
    &=
    \sum_{i=1}^{m}
    \left(
        \prod_{k=1}^{i-1} \detul{(1-\alpha_k)} \, \attul{\alpha_i}
    +
        \sum_{j=1}^{i-1}
        \prod_{\substack{k=1 \\ k\neq j}}^{i-1}
        \detul{(1-\alpha_k)} \,
        (1 - \attul{\alpha_j}) \,
        \detul{\alpha_i}
    \right) \,
    \ell(L_i)
    \notag \\
    &=
    \sum_{i=1}^{m}
        \left(
            \prod_{k=1}^{i-1} (1 - \attul{\alpha_k}) \, \attul{\alpha_i} \, \detul{\ell(L_i)} \,
        \right) + c_1,
\end{align}
where $c_1$ is a constant value.
Finally, we can insert this result back in \autoref{eq:loss_ray2} to obtain a loss where all variables can be differentiated:
\begin{align}
    & \argmin ~ \mathrm{(a)} \!+\! \mathrm{(b)} \!+\! \mathrm{(c)}
    \notag \\
    \overset{\eqref{eq:loss_c3}}{=}
    & \argmin
    \sum_{i=1}^{m}
        \left(
            \prod_{k=1}^{i-1} (1 - \attul{\alpha_k}) \, \attul{\alpha_i} \, \detul{\ell(L_i)} \,
    +
            \prod_{k=1}^{i-1} \detul{(1-\alpha_k) \, \alpha_i} \, \ell(\attul{L_i}) \,
        \right)
    \notag \\
    = & \argmin
    \sum_{i=1}^{m}
        \left(
            \prod_{k=1}^{i-1} (1 - \attul{\alpha_k}) \, \attul{\alpha_i} \, \ell(\attul{L_i})
        \right).
\end{align}
This final result is equivalent to the one shown in \autoref{fig:nerf_comparison} of the main document.
It is now also apparent that we do not need to produce extra samples to evaluate $\E_{t>t_\p}[\ell(L_t)]$.

\section{Design space of the background distribution}
\label{sec:choice_of_distribution}

\begin{figure}[t]
    \centering
    \includegraphics[width=\linewidth]{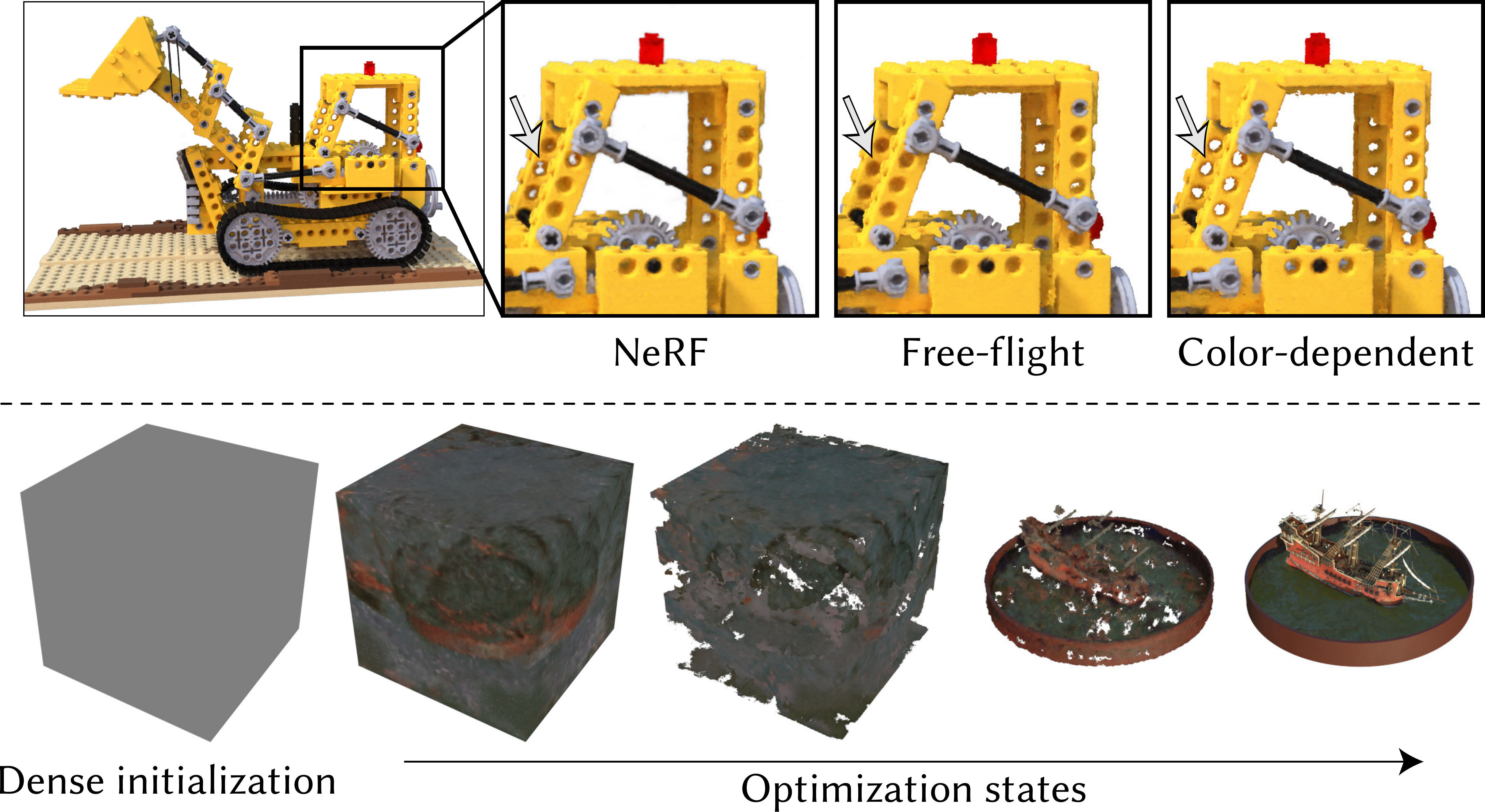}
    \caption{\label{fig:VDT}%
        \textbf{Benefits of the color-dependent background distribution.}
        (Top) After $2000$ iterations, the color-dependent variant explores further along the ray and clears over aggressively optimized regions faster.
        (Bottom) In a contrived experiment where the scene is densely initialized, the optimizer first attempts to bake the images onto the cube.
        The color-dependent variant can penetrate high-occupancy regions, while others get stuck.
        All experiments are conducted with the same hyperparameters in the Instant NGP \citep{mueller2022instant} codebase.
    }
\end{figure}

\autoref{sec:stochastic_baseline} of the main document proposes the stochastic background surface.
Designing the background surface distribution $\fb$ (\autoref{fig:appendix_bg}) involves a tradeoff between exploitation and exploration, offering a wide design space.

On one hand, choosing a background surface close to the model's current best guess (i.e., in high occupancy regions) ensures that only perturbations with an improvement will be accepted.
An example is the deterministic strategy to always use the $0.5$ level set.
It aggressively selects the first potential surface with more than $50\%$ confidence along the ray as the background, ignoring any further possibilities.

On the other hand, exploring more possibilities enhances the algorithm's robustness in scenes with complex occlusions.
The free-flight background distribution is a softer version of the deterministic strategy.
Instead of using a threshold to binarize the occupancy field, it stochastically decides whether to use a surface as the perturbation background during ray traversal.

Many other designs are possible.
One unique to our method is the \emph{color-dependent background distribution}.
Unlike the free-flight distribution, which relies only on the occupancy value, this approach also considers how well each potential surface aligns with the target color, measured by $\ell(L_\p, \, L_\mathrm{target})$.
The additional information enables us to discard high-occupancy surfaces that poorly match the target color, which may result from overly aggressive optimization.
Specifically, we compute an effective occupancy $\alpha'$, as a modification of the original occupancy $\alpha$:
\begin{align*}
    \alpha' = \frac{\alpha}{1 + c \, \ell(L_\p, L_\mathrm{target})}.
\end{align*}
When the color matches well, the transformation is neutral, but for misaligned colors, it reduces the effective occupancy, lowering the likelihood of selecting such surfaces as a background. In our experiments, we used $c=16$.
As shown in \autoref{fig:VDT}, this color-dependent distribution can be more efficient when penetrating incorrect surfaces.

This expanded design space is particularly compelling.
Traditional reconstruction methods mostly optimize in image space, interacting with the 3D scene only through a rendering algorithm (surface-based or volume-based).
As a result, the design space in image space is quite limited, with little to do beyond computing a loss.

In contrast, our method operates directly in the scene space.
Background distributions can be tailored to focus on regions of interest.
This flexibility enables the development of new optimization strategies that are unattainable in image-space methods.
The color-dependent background distribution is an example to actively guide the optimization to skip regions that are \emph{believed} to be wrong regardless of the occupancy value.

In this paper, we focus on the free-flight distribution to highlight the dual-loss relationship with NeRF, leaving the exploration of other distributions for future work. Only the result in \autoref{fig:VDT} uses the color-dependent distribution.

\section{Additional experiments and results}
\label{sec:additional_experiments_and_results}

\paragraph{Interior topology changes}

Methods based on local surface evolution struggle with interior topological changes, like transforming a sphere into a torus.
Indeed, they primarily rely on deforming visibility silhouettes to change the overall shape, but these silhouettes often do not exist in regions away from the outer contour.

Correctly handling such topological changes requires making a significant modification, such as cutting a cone through the entire object to expose the occluded background.
This type of change is beyond the reach of common derivative-based methods, which can only account for infinitesimal perturbations.
\begin{center}
    \vspace{2mm}
    \includegraphics[width=0.70\columnwidth]{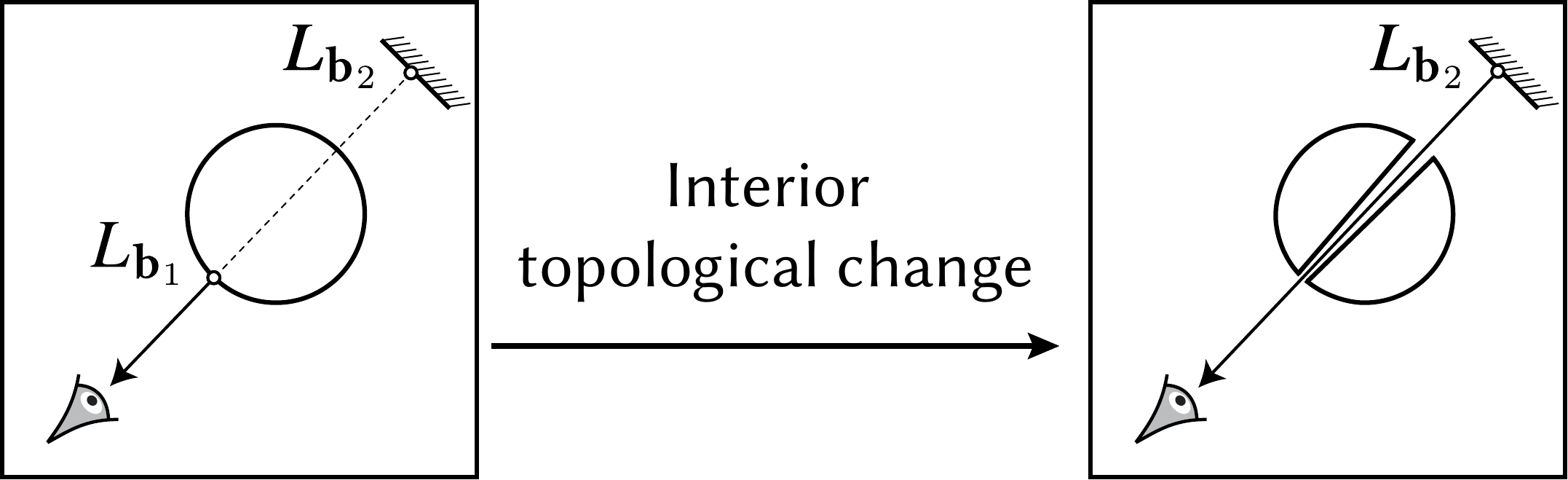}
    \vspace{0mm}
\end{center}

\citeauthor{mehta2023theory}~\shortcite{mehta2023theory} propose a cone-shaped perturbation strategy to test whether exposing the background improves the match to the target color in the application of physically based rendering. This approach significantly improves convergence in scenes that require hole penetration, compared to conventional surface evolution methods.

\begin{figure}[t]
    \centering
    \includegraphics[width=\linewidth]{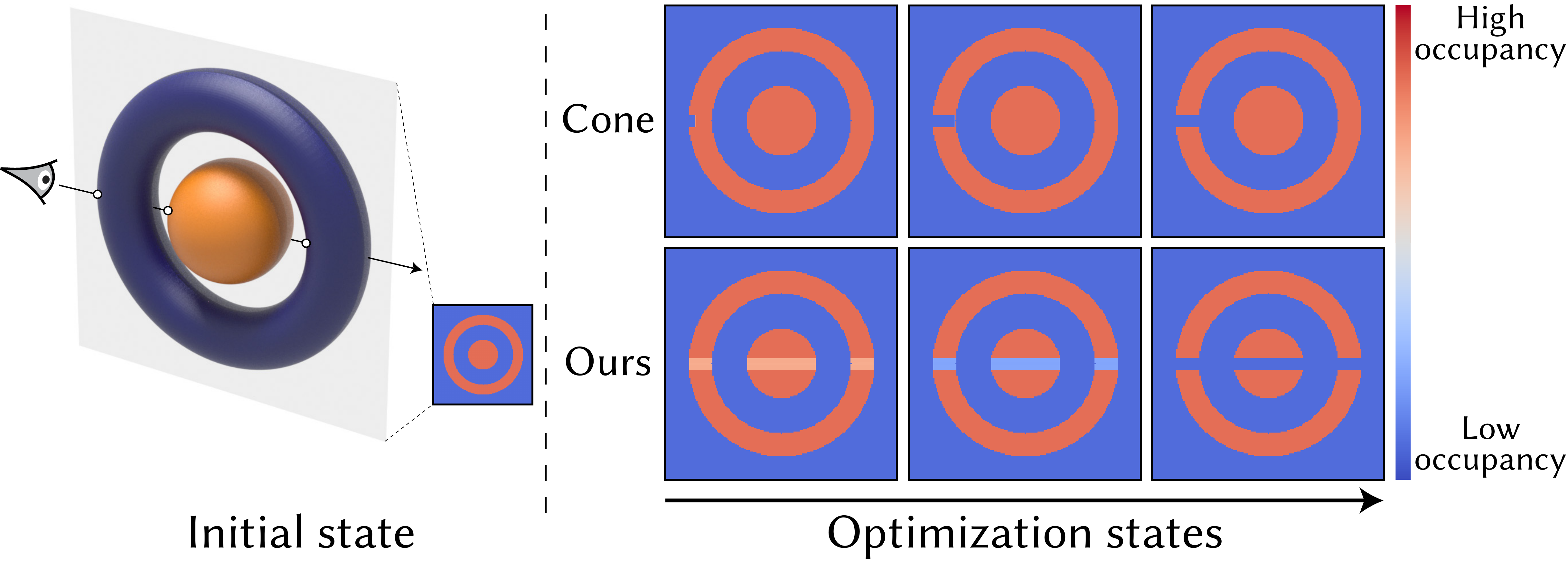}
    \caption{\label{fig:cone_perturbation}%
        \textbf{Left: } We test interior topological changes in a scene where orange better aligns with the target background color than indigo.
        \textbf{Right: } We show optimization states by visualizing a 2D slice of the occupancy field.
        The cone perturbation strategy \citep{mehta2023theory} gets stuck after penetrating the torus once, as it can only see through a single obstacle.
    }
\end{figure}

However, this strategy can also affect scenes where the topology is already correct. In such cases, only local refinement is needed, and the cone perturbation may bias the derivative in the wrong direction. Additionally, the cone perturbation strategy can only penetrate a single obstacle, limiting its ability to handle complex real-world scenes that require penetration through multiple layers of geometry (\autoref{fig:cone_perturbation}).
Our stochastic background strategy addresses these challenges by considering additional background possibilities, enabling more robust optimization for complex scenes.

\begin{figure*}[h]
    \centering
    \includegraphics[width=\textwidth]{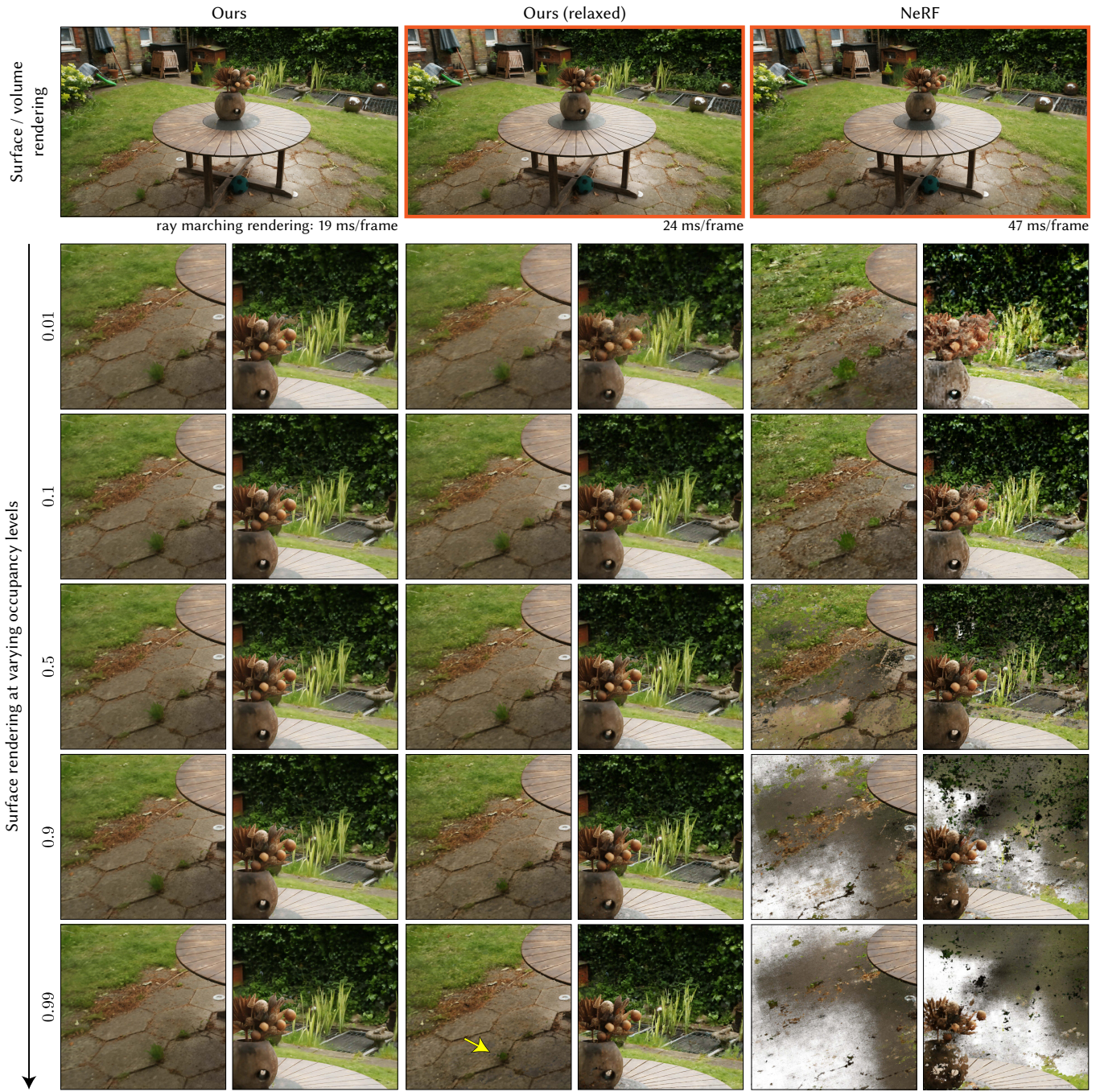}
    \caption{Surface rendering at varying level sets of a scene reconstructed by our method and NeRF, using the same hyperparameters.
    Only the two images with orange borders are rendered volumetrically.
    \textbf{Left:} For our method, the surface rendering shows minimal changes across different level set thresholds, indicating that the occupancy field has converged to a near-Heaviside step function on the surface.
    \textbf{Middle:} The relaxed variant of our algorithm uses volume representation in challenging regions, such as sub-pixel details (yellow arrow). The overall scene remains surface-like, leading to better ray marching performance than NeRF.
    \textbf{Right:} The NeRF reconstruction is inherently volumetric, thus renderings of level sets do not produce meaningful visualizations.}
    \label{fig:level_sets2}
\end{figure*}

\paragraph{Rendering}
Once the occupancy field trained with our algorithm has converged, it should have value $0$ in empty space and $1$ on the surface. Since our field storage is continuous in practice, we aim for a near Heaviside step function on the surface. In \autoref{fig:level_sets2}, we show an additional level set rendering result for an outdoor scene to demonstrate that our method can achieve this, with any level set being usable. In this paper we use $0.5$ as the threshold.

We propose two methods for rendering the level set. The first method involves ray marching with a small step size. In this approach, we immediately return the color of the first sample point that hits the surface (when occupancy exceeds $0.5$), without any weighting or color blending. The second method involves extracting a triangle mesh using marching cubes or TSDF fusion, then rasterizing the mesh to obtain the hit point location and querying the color network for the final color.

\begin{figure}[t]
    \centering
    \includegraphics[width=\linewidth]{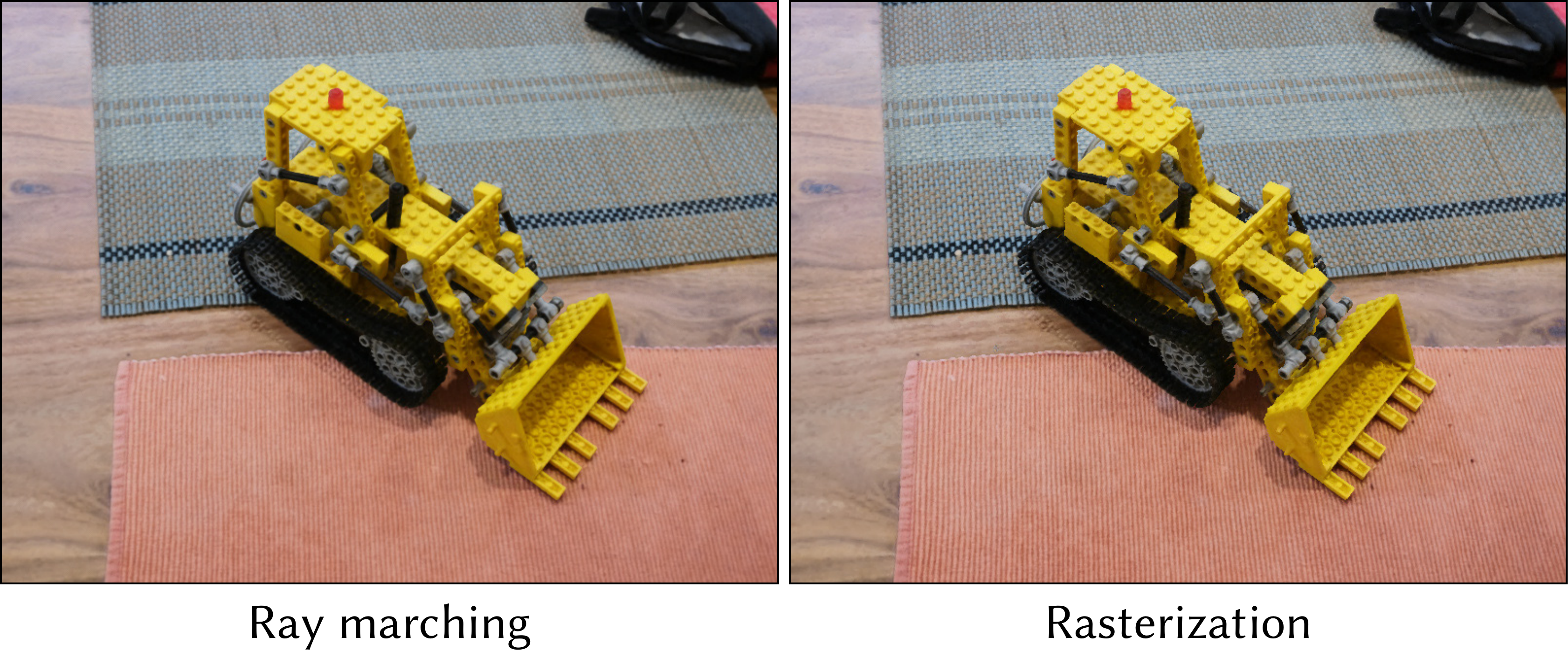}
    \caption{\label{fig:rasterization}%
    Visual comparison of the same surface scene trained with our algorithm using two rendering methods: ray marching (left) and mesh rasterization (right). Both methods give nearly identical results.
    }
\end{figure}

Both methods produce nearly identical visual results, as shown in \autoref{fig:rasterization}.

\begin{figure}[t]
    \centering
    \includegraphics[width=1.0\linewidth]{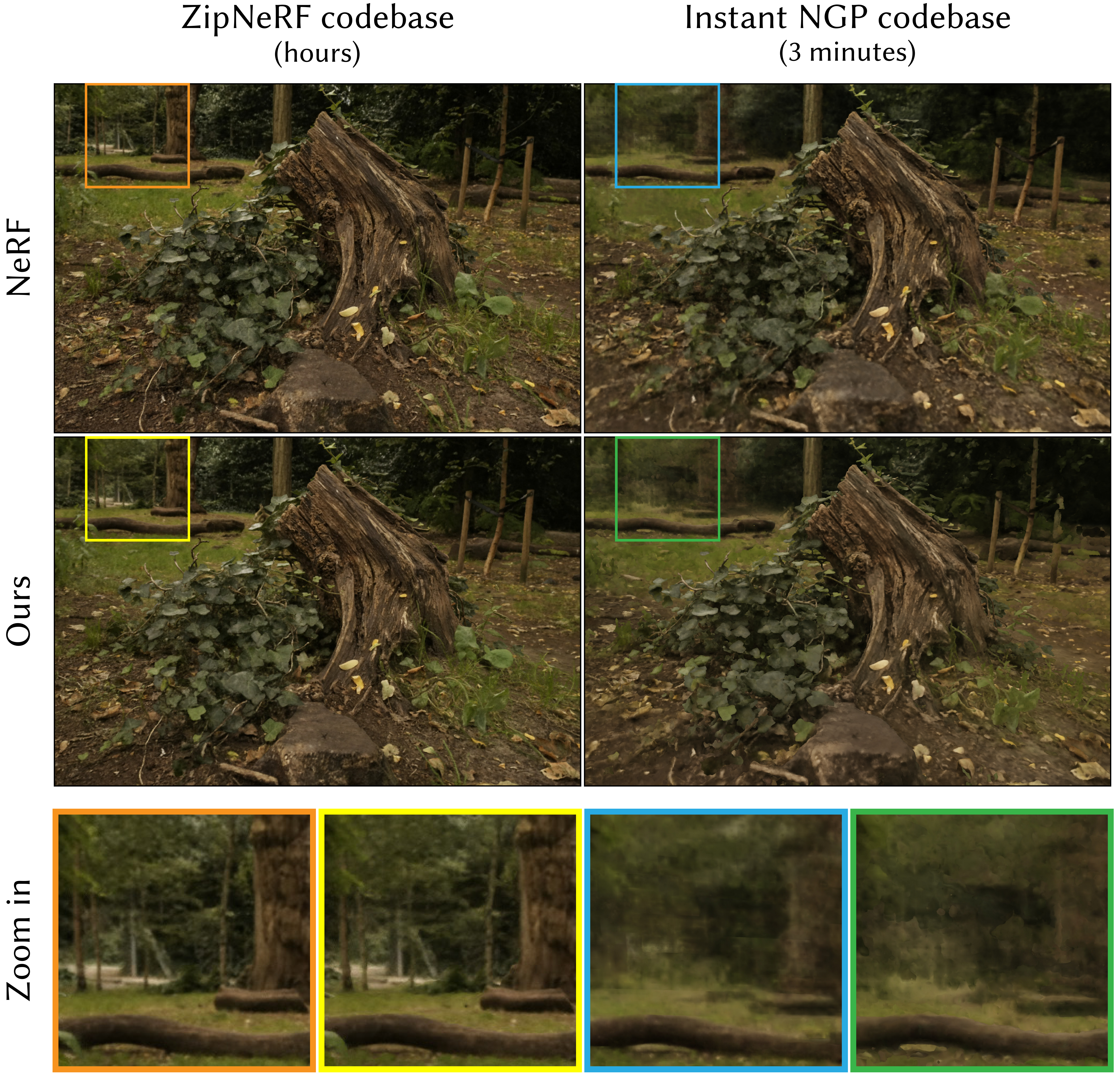}
    \caption{
        Qualitative comparison of NeRF and our method in two codebases.
    }\label{fig:zipnerf}
\end{figure}

\paragraph{Codebase}
Our work primarily focuses on the theoretical development of a surface-based scene reconstruction algorithm, while the specifics of the model implementation are largely independent of our core algorithm.
For example, the Instant NGP codebase is optimized for speed and designed for object-centric scenes, resulting in suboptimal details in the far field background (\autoref{fig:zipnerf}).
Our results inherit these advantages and limitations.

\begin{figure}[t]
    \centering
    \includegraphics[width=\linewidth]{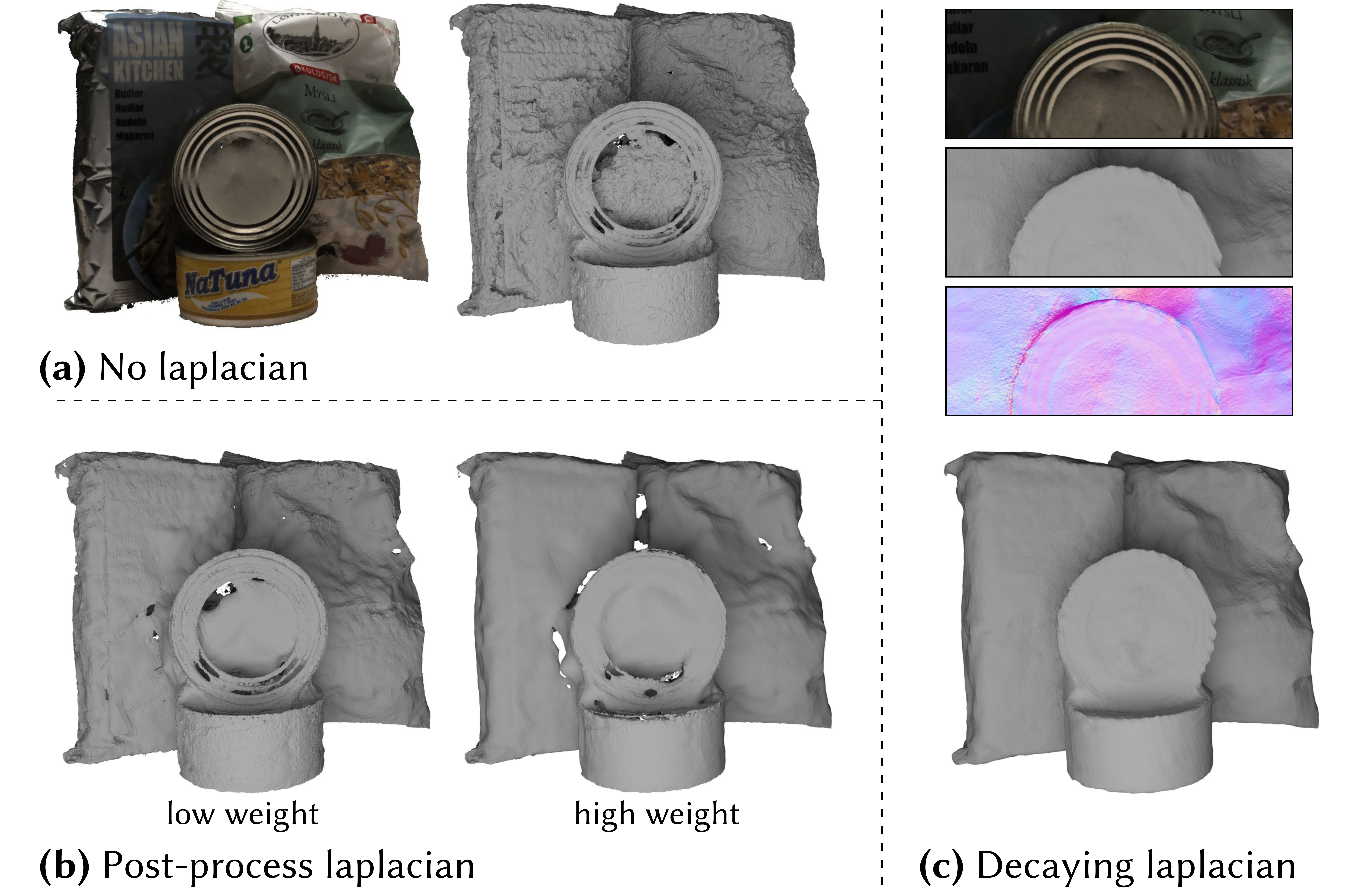}
    \caption{\label{fig:exponential_decay}%
        For highly underconstrained scenes with shiny surfaces and limited viewing angles, training our algorithm with an exponentially decaying Laplacian is more effective than applying Laplacian as a post-processing step.
    }
\end{figure}

\paragraph{Decaying Laplacian}
For simple scenes with sufficient observations, Laplacian smoothing as a post-processing step can effectively refine surface geometry.
However, this approach has limitations in more challenging scenarios.
As shown in \autoref{fig:exponential_decay}, we analyze a highly underconstrained scene with shiny surfaces that exhibit rapid color changes with viewing angle, captured only from the front.
Here, training without Laplacian smoothing achieves good novel view synthesis but results in geometry errors, particularly at the can's bottom.

Applying a Laplacian as a post-processing step requires numerous iterations to address these issues and may degrade geometry in other regions.
In contrast, training our algorithm with an exponentially decaying Laplacian is more efficient.
Consequently, the results in \autoref{fig:geometry_big} of the main document are obtained by training our algorithm with an exponentially decaying Laplacian.

\begin{figure}[t]
    \centering
    \includegraphics[width=\linewidth]{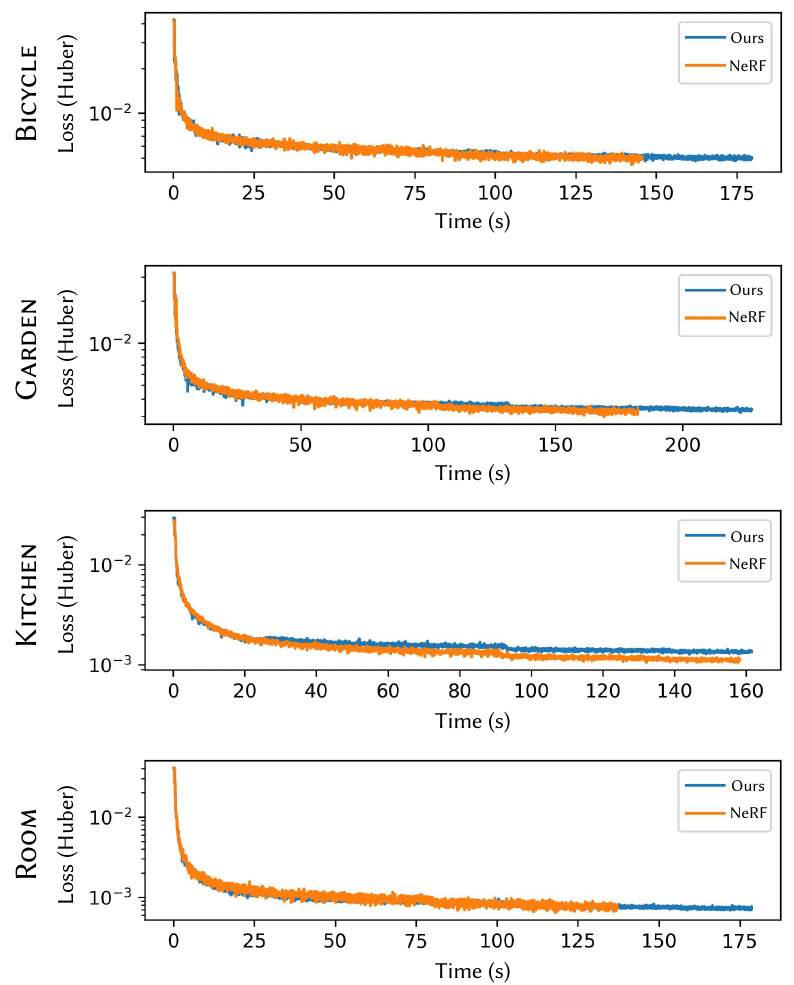}
    \caption{\label{fig:optimization_curve}%
        \textbf{Equal time convergence plot.} Our method converges at a rate comparable to NeRF in the Instant NGP codebase. We have a longer tail in the loss curve since our method spawns more rays per iteration than NeRF.
    }
\end{figure}

\paragraph{Training time}
\autoref{fig:optimization_curve} shows the loss convergence plot in the Instant NGP codebase, demonstrating that our method converges at a rate comparable to NeRF despite its surface reconstruction nature.

Like NeRF, our method computes the loss in linear time only using per-sample occupancy and color values along a ray.
Theoretically, this ensures it is at least as fast as NeRF.
However, the observed increase in training time arises from INGP's training strategy, which targets a fixed sample size per iteration (we use the default value $2^{18}$) by spawning as many rays as needed.
Since our method reconstructs surfaces, it typically requires fewer samples along rays in near-converged regions, allowing more rays to be processed within the same sample budget.
In practice, the increase in ray count causes INGP to become slower.

This slowdown is a consequence of INGP's implementation rather than a limitation of our method.
In fact, our method's efficiency in using fewer resources per ray is advantageous.
This also explains why our relaxed variant achieves a higher PSNR than NeRF: it utilizes the same sample budget to visit more reference pixels per iteration.

\paragraph{Miscellaneous}
\autoref{fig:level_sets2} shows additional level set rendering results for an outdoor scene.
\autoref{tab:psnr_comparison}, \autoref{tab:ssim_comparison} and \autoref{tab:lpips_comparison} show the complete PSNR, SSIM and LPIPS results in the Instant NGP codebase.
\autoref{tab:psnr_comparison_zipnerf} shows the complete PSNR results in the ZipNeRF codebase.
\autoref{tab:cd} shows the complete Chamfer distance results on the DTU dataset.

\section{Volume relaxation}
This section details a heuristic-based volume relaxation of our method.
While we do not claim this to be the only way to relax our method, it provides a straightforward and effective way to retain the surface-like properties of the scene while enabling volumetric blending in regions where the surface representation is insufficient.

We propose the following loss function as a relaxed volumetric version of our loss in the form of \autoref{eq:loss_ray}.
The notation $\attul{(\cdot)}$ is consistent with \autoref{sec:expectation_over_bg}, denoting terms that are differentiated during optimization:
\begin{align}
    \label{eq:relaxed_loss_ray}
    \loss_\mathrm{ray}^{\mathrm{vol}}
    =
    \sum_{i=1}^{m}
    \left[
    \left(  \prod_{k=1}^{i - 1} (1 - \alpha_k)   \right) \,
    \ell
    \Big(
        \attul{\alpha_i} \, \attul{L_i}
        + (1 - \attul{\alpha_i}) \, \E_{t>t_i}[L_t]
        , \,
        L_\mathrm{goal}
    \Big)
    \right],
\end{align}
where the error metric $\ell$ now compares against a modified target color $L_\mathrm{goal}$:
\begin{align}
    L_\mathrm{goal}
    =
    \frac{L_\mathrm{target} - L_\mathrm{prev}}{T_\mathrm{prev}}
    =
    \frac{
        L_\mathrm{target} - \sum_{j=1}^{i-1}  \left(\prod_{k=1}^{j-1} (1 - \alpha_k) \right)  \alpha_j  L_j
    }{\prod_{j=1}^{i-1} (1 - \alpha_j)}.
\end{align}
\autoref{eq:relaxed_loss_ray} is derived from two key modifications to the radiance field loss (\autoref{eq:loss_ray}):
\begin{itemize}
    \item We now blend colors instead of error metrics to allow for volumetric blending for the $i$-th sample.
    \item The $i$-th sample no longer needs to match the target color $L_\mathrm{target}$ directly. Instead, its goal adjusts for the color contribution of prior samples $L_\mathrm{prev}$ and the transmittance from the camera to the $i$-th sample $T_\mathrm{prev}$.
\end{itemize}
Empirically, this relaxed loss performs well as a volume reconstruction algorithm.
However, when used to refine a converged surface scene, this loss often converts the entire scene into a volumetric representation, even in regions where the surface representation is already visually adequate.
This happens because a surface representation is essentially a special case of a volume with fewer degrees of freedom, and fitting colors in a volume generally reduces the loss more easily than fitting colors on a surface.

To prevent over-relaxation, we propose a heuristic to detect locations where volume relaxation is unnecessary.
Specifically, when the local loss without blending at a specific position is no worse than the local loss with blending:
\begin{align}
    \ell(L_i, L_\mathrm{goal})
    \leq
    \ell
    \left(
    \alpha_i \, L_i
    + (1 - \alpha_i) \, \E_{t>t_i}[L_t]
    , \,
    L_\mathrm{goal}
    \right),
\end{align}
we use the local loss without blending in \autoref{eq:relaxed_loss_ray}.
This comparison does not introduce any overhead, as all necessary values are already available.
Our experimental results show that this heuristic is effective in preserving Heaviside-like occupancy values in most areas while allowing for volumetric blending in challenging regions (\autoref{fig:level_sets2}).

We highlight again that the volume relaxation step is a heuristic and not a fundamental part of our method.
All results are obtained without this relaxation in this paper unless explicitly stated.

\section{Implementation details}

All results were generated and measured on a Linux workstation with an AMD Ryzen 7950X processor and an NVIDIA RTX 4090 graphics card.

\paragraph{Instant NGP codebase}
We used the default hyperparameter configuration file (base.json) provided by the authors and retained the original sampling strategy. However, we made two key modifications to the codebase to accommodate our method:
\begin{itemize}
    \item We reduced the ray marching step size from $1/1024$ to $1/2048$ to achieve a finer surface resolution.
    \item The maximum buffer size for storing temporary samples was increased from $16 \times \mathrm{target~batch~size}$ to $128 \times \mathrm{target~batch~size}$ to accommodate the increased number of rays spawned in each iteration.
\end{itemize}

Since INGP does not natively support automatic differentiation, we manually implemented the derivative propagation of our method into the codebase, similar to how the framework trains NeRF.

For the geometry reconstruction experiments shown in \autoref{fig:geometry_big} (main document), we used a $L_1$ loss to improve convergence in dark regions. Models were trained for $10000$ iterations (reduced from the default $35000$), with the Laplacian weight decaying exponentially to $2 \times 10^{-5}$. The Laplacian was estimated via finite differences using six neighboring samples with an epsilon of $1/1024$ (approximately 1 mm for a unit cube).

Rendering times were measured without DLSS.

\paragraph{ZipNeRF codebase}
We used the default hyperparameter configuration file (360.gin) along with the original adaptive sampling strategy. As ZipNeRF’s adaptive sampling is tailored for volume reconstruction, it may not be optimal for our method. However, we deliberately avoided modifying these components to minimize intrusive changes and focus on proof-of-concept validation.

\paragraph{Warm start}
During training, our algorithm can sometimes push occupancy values in certain regions (e.g., peripheral or camera-adjacent areas) too high in early stages, resulting in floaters in the final reconstruction.
This occurs because the background is insufficiently explored at the beginning, leading to overly aggressive optimization of temporarily superior candidates.
While NeRF encounters similar issues, recovery is particularly challenging in our case since occupancy values of these floaters could approach $1$.

For INGP training, we can mitigate this issue by adjusting the learning rate schedule at the cost of slower convergence.
Empirically, we found it also effective to impose a moving upper bound on occupancy values, gradually relaxing this constraint during training.
Specifically, at iteration $i$, we bound the occupancy value by $\alpha_\mathrm{max} = 0.1 + 0.9 \times i / 1000$.
This constraint is active only during the first $1000$ iterations, corresponding to the first few seconds of training.
Additionally, we observed that our relaxed training strategy is less prone to floaters.
For novel view synthesis tasks, we trained the relaxed variant for $5000$ iterations as a warm start.

The floater issue also pops up in the ZipNeRF codebase.
For simplicity, we adopted a NeRF training warm start during the first
5\% of training iterations and did not bound occupancy values.

\begin{table*}[t]
    \caption{ PSNR comparison using the Instant NGP codebase. Ours uses surface rendering, while the relaxed variant and NeRF use volume rendering.}
    \vspace{-2mm}
    \label{tab:psnr_comparison}
    \begin{tabular}{llllllllll}
    \toprule
                   & Bicycle & Bonsai & Counter & Garden & Kitchen & Room  & Stump & Flowers & Treehill \\
    \midrule
    Ours           & 22.53   & 31.22  & 26.67   & 23.81  & 28.58   & 29.59 & 24.16 & 19.76   & 21.79    \\
    Ours (relaxed) & 22.66   & 31.81  & 26.95   & 24.04  & 29.14   & 29.75 & 24.43 & 19.98   & 21.97    \\
    NeRF           & 22.66   & 31.45  & 26.79   & 23.97  & 29.33   & 29.17 & 23.96 & 19.95   & 21.82    \\
    \bottomrule
    \end{tabular}
\end{table*}

\begin{table*}[t]
    \caption{SSIM comparison using the Instant NGP codebase.}
    \vspace{-2mm}
    \label{tab:ssim_comparison}
    \begin{tabular}{llllllllll}
    \toprule
                   & Bicycle & Bonsai & Counter & Garden & Kitchen & Room  & Stump & Flowers & Treehill \\
    \midrule
    Ours           & 0.673 & 0.918 & 0.872 & 0.686 & 0.866 & 0.896 & 0.769 & 0.577 & 0.692    \\
    Ours (relaxed) & 0.682 & 0.927 & 0.882 & 0.695 & 0.878 & 0.902 & 0.784 & 0.590 & 0.698    \\
    NeRF           & 0.675 & 0.924 & 0.877 & 0.687 & 0.877 & 0.893 & 0.776 & 0.586 & 0.692 \\
    \bottomrule
    \end{tabular}
\end{table*}

\begin{table*}[t]
    \caption{LPIPS comparison using the Instant NGP codebase.}
    \vspace{-2mm}
    \label{tab:lpips_comparison}
    \begin{tabular}{llllllllll}
    \toprule
                   & Bicycle & Bonsai & Counter & Garden & Kitchen & Room  & Stump & Flowers & Treehill \\
    \midrule
    Ours           & 0.578 & 0.241 & 0.315 & 0.547 & 0.236 & 0.306 & 0.475 & 0.618 & 0.599    \\
    Ours (relaxed) & 0.642 & 0.244 & 0.335 & 0.672 & 0.234 & 0.324 & 0.497 & 0.676 & 0.645    \\
    NeRF           & 0.658 & 0.256 & 0.354 & 0.625 & 0.239 & 0.362 & 0.514 & 0.699 & 0.692 \\
    \bottomrule
    \end{tabular}
\end{table*}

\begin{table*}[t]
    \caption{PSNR comparison using the ZipNeRF codebase.}
    \vspace{-2mm}
    \label{tab:psnr_comparison_zipnerf}
    \begin{tabular}{llllllllll}
    \toprule
         & Bicycle & Bonsai & Counter & Garden & Kitchen & Room  & Stump & Flowers & Treehill \\
    \midrule
    Ours & 24.10   & 31.24  & 26.38   & 26.14  & 30.22   & 31.07 & 25.96 & 20.99   & 23.12    \\
    NeRF & 25.50   & 33.20  & 28.16   & 27.62  & 32.01   & 32.44 & 27.11 & 22.11   & 23.85    \\
    \bottomrule
    \end{tabular}
\end{table*}

\begin{table*}[t]
    \caption{Chamfer Distance comparison on the DTU dataset
     with NeuS \cite{wang2021neus} and NeuS2 \cite{wang2023neus2}.}
    \vspace{-2mm}
    \label{tab:cd}
    \begin{tabular}{lllllllll}
    \toprule
         & Scan24 & Scan37 & Scan40 & Scan55 & Scan63 & Scan65 & Scan69 & Scan83 \\
    \midrule
    Ours (1 minute) & 0.81 & 0.77 & 0.66 & 0.40 & 1.08 & 0.90 & 0.88 & 1.42    \\
    NeuS (8 hours) & 0.83 & 0.98 & 0.56 & 0.37 & 1.13 & 0.59 & 0.60 & 1.45    \\
    NeuS2 (5 minutes) & 0.56 & 0.76 & 0.49 & 0.37 & 0.92 & 0.71 & 0.76 & 1.22   \\
    \midrule
        & Scan97 & Scan105 & Scan106 & Scan110 & Scan114 & Scan118 & Scan122 & \\
        \midrule
    Ours & 1.20 & 0.75 & 0.68 & 1.07 & 0.61 & 0.55 & 0.63 & \\
    NeuS & 0.95 & 0.78 & 0.52 & 1.43 & 0.36 & 0.45 & 0.45 & \\
    NeuS2 & 1.08 & 0.63 & 0.59 & 0.89 & 0.40 & 0.48 & 0.55 & \\
    \bottomrule
    \end{tabular}
\end{table*}

\end{document}